\documentclass{article}
\usepackage[utf8]{inputenc}
\usepackage[margin=1in]{geometry}
\usepackage{subcaption}
\usepackage{graphicx}
\usepackage{amsmath}
\usepackage[hyphens]{url}
\usepackage{hyperref}
\hypersetup{colorlinks=false,breaklinks=true,hidelinks}
\usepackage{cleveref}
\usepackage{apacite}
\usepackage{natbib}
\usepackage{authblk}
\usepackage{multirow}
\graphicspath{{figures/}}

\DeclareMathOperator*{\minimize}{minimize}

\title{Earthquake Phase Association using a \\ Bayesian Gaussian Mixture Model}
\author[1]{Weiqiang Zhu}
\author[1]{Ian W. McBrearty}
\author[1]{S. Mostafa Mousavi}
\author[1]{\\ William L. Ellsworth}
\author[1]{Gregory C. Beroza}
\affil[1]{\small Department of Geophysics, Stanford University, Stanford, CA, 94305}
\date{}

\begin{document}

\maketitle

\begin{abstract}
Earthquake phase association algorithms aggregate picked seismic phases from a network of seismometers into individual earthquakes and play an important role in earthquake monitoring. Dense seismic networks and improved phase picking methods produce massive earthquake phase data sets, particularly for earthquake swarms and aftershocks occurring closely in time and space, making phase association a challenging problem. We present a new association method, the \textit{Ga}ussian \textit{M}ixture \textit{M}odel \textit{A}ssociation (GaMMA), that combines the Gaussian mixture model for phase measurements (both time and amplitude), with earthquake location, origin time, and magnitude estimation. We treat earthquake phase association as an unsupervised clustering problem in a probabilistic framework, where each earthquake corresponds to a cluster of P and S phases with hyperbolic moveout of arrival times and a decay of amplitude with distance. We use a multivariate Gaussian distribution to model the collection of phase picks for an event, the mean of which is given by the predicted arrival time and amplitude from the causative event. We carry out the pick assignment for each earthquake and determine earthquake parameters (i.e., earthquake location, origin time, and magnitude) under the maximum likelihood criterion using the Expectation-Maximization (EM) algorithm. The GaMMA method does not require the typical association steps of other algorithms, such as grid-search or supervised training. The results on both synthetic test and the 2019 Ridgecrest earthquake sequence show that GaMMA effectively associates phases from a temporally and spatially dense earthquake sequence while producing useful estimates of earthquake location and magnitude. 

\end{abstract}

\section{Introduction}

Earthquake catalogs are fundamental products that are widely used in seismology to study and model various aspects of seismicity. Extensive efforts have been made to generate more complete catalogs with many more smaller earthquakes and more precise location and magnitude estimates. These high-resolution high-precision catalogs have the potential to reveal relationships among earthquakes and illuminate active structures that would otherwise remain hidden \citep{waldhauser2008large, hauksson2012waveform, yoon2015earthquake, ross2019searching, park2020machine, tan2021machine, beroza2021machine}.
A standard earthquake-monitoring workflow from seismic waveforms to earthquake catalogs includes several tasks, including earthquake detection, phase picking \citep{allen1978automatic}, phase association \citep{yeck2019glass3}, earthquake location \citep{klein2002user}, and magnitude estimation \citep{richter1935instrumental}.

The phase picking step detects seismic phases such as P-wave and S-wave phases at each seismic station. The phase association step aggregates these phases from multiple stations of a seismic network into separate groups associated with each earthquake. Earthquake location and magnitude are then estimated from the associated phase information, i.e., arrival time and amplitude. The resulting catalog can be further enhanced through template matching \citep{gibbons2006detection, shelly2007non, peng2009migration} or subspace projection \citep{harris2011, barrett2014} that use the detected earthquakes as templates to re-scan the waveforms and detect small earthquakes with similar waveforms.

The phase picking step has been significantly improved by deep-learning-based pickers that learn from manual picks labeled by analysts to detect millions of phase picks from raw seismic waveforms \citep{ross2018generalized, zhu2019phasenet, mousavi2020earthquake}. The rapidly growing volume of automatic picks and the ongoing growth of seismic networks makes developing effective phase association methods crucial. Phase association has not yet received the attention that has been devoted to other earthquake monitoring tasks. Association methods based on back-projection are most commonly used in classic earthquake monitoring systems, such as GLASS3 \citep{yeck2019glass3}, Earthworm \citep{friberg2010earthworm}, and SeisComP3 \citep{weber2007seiscomp3}. These approaches usually deploy a grid-search and back-project phase picks based on the expected moveout with distance. An earthquake and its initial location is declared based on the number of phase picks inside a spatial grid that are consistent with a candidate location. Although back-projection-based association is robust and effective, its performance is limited for dense earthquake sequences when earthquakes occur so closely in time and space that interpreting the maxima resulting from back-projection becomes problematic. 
Studies have continued to focus on improving the grid-search and back-projection approach for different scenarios \citep{draelos2015new, gibbons2016iterative, zhang2019rapid}.
Meanwhile, several new approaches have been proposed to solve the earthquake phase association using: graph theory \citep{mcbrearty2019earthquake}, the RANSAC algorithm \citep{woollam2020hex, zhu2021multi}, and deep learning \citep{ross2019phaselink, mcbrearty2019pairwise, dickey2020beyond}.
When combined with the rapid development of phase picking methods, better association methods have the potential to improve significantly the overall performance of earthquake monitoring pipelines. 

We propose an association method based on a Bayesian Gaussian mixture model, which is an unsupervised machine learning method for clustering \citep{bishop2006pattern} that has been widely used in different research fields, such as image processing \citep{permuter2006study}, speech recognition \citep{reynolds1995robust}, and earthquake studies \citep{ross2020directivity, seydoux2020clustering}. 
Earthquake phase association can be treated as an unsupervised clustering problem, with groups of phase picks, in time and space, arising from a discrete set of earthquake origins. We combine the Gaussian mixture model with earthquake location, origin time, and magnitude estimation, so that the GaMMA method can cluster phase picks based on the physical constraints of arrival time moveout and amplitude decay with distance. Phase amplitude information is often neglected because it can be difficult to account for in conventional association methods; however, GaMMA is designed such that it can use phase arrival time, phase-type identification, and amplitude information for association while simultaneously estimating the underlying event source characteristics (i.e., location and magnitude). Moreover, GaMMA does not require extra association steps of grid-search or supervised training. These attributes make GaMMA an appealing approach to address the challenges arising in processing of large numbers of automatic picks in earthquake monitoring workflows.

\section{Method}
The objective of earthquake phase association is to post-process a large collection of phases picked on individual seismic stations and cluster them into groups of seismic phases originating from a same earthquake event, so that subsequent earthquake characterization tasks can be performed on individual events.
Phase arrival times from the same earthquake follow a hyperbolic moveout that is determined by the hypocentral distance and the Earth model (i.e. seismic wave speed). This moveout allows association algorithms to distinguish between phases from different earthquakes.
In this work, we extend the association problem by using both phase arrival time and phase amplitude information. On average, the phase amplitude scales with earthquake magnitude and decays with the hypocentral distance. Thus, the amplitude provides additional information to improve phase association. We formulate the association problem as follows: Given $N$ seismic phases $(x_i, y_i, z_i, t_{i}, a_{i})$, i.e., arrival time $t_i$ and amplitude recorded $a_i$ at the $i$-th seismic station located at $(x_i, y_i, z_i)$, we seek to group these phases into $K$ earthquakes and estimate the underlying source parameters $(x_k, y_k, z_k, t_k, m_k)$, i.e., location $(x_k, y_k, z_k)$, origin time $t_k$, and magnitude $m_k$ of the $k$-th earthquake.

We solve this association problem using the Gaussian mixture model \citep{permuter2006study}, which is an unsupervised clustering method that groups $N$ data points into $K$ clusters by maximizing the probability that these $N$ data can be explained by a mixture of $K$ Gaussian distributions.

\begin{figure}
    \centering
    \includegraphics[width=0.8\textwidth]{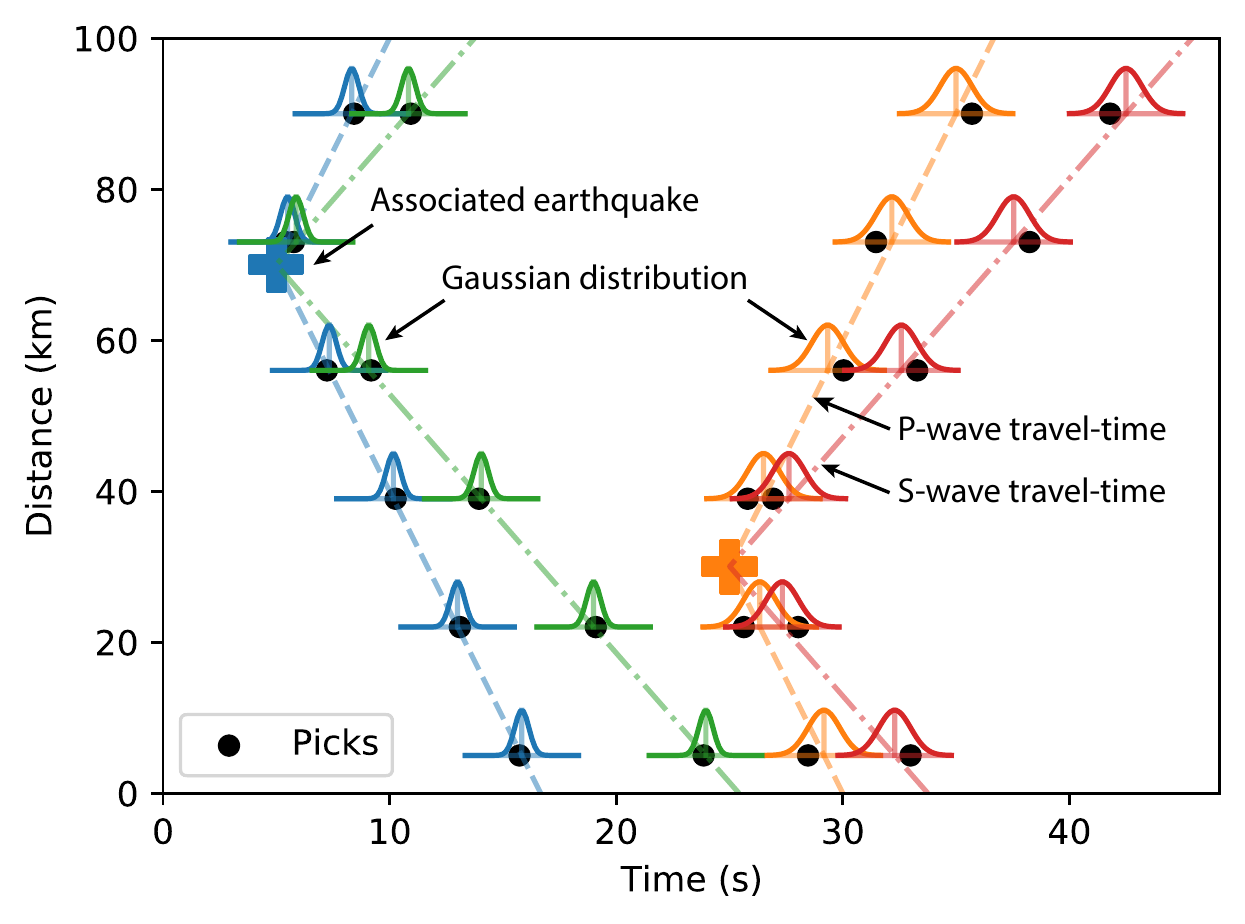}
    \caption{Gaussian Mixture Model for Association. The GaMMA method models Gaussian distributions based on the theoretical phase travel-time and amplitude and uses the Expectation-Maximization (EM) algorithm to update iteratively: phase assignment, earthquake source parameters, and the mean and standard deviation of Gaussian distribution. This iteration converges to the correct phase assignment and earthquake source parameters to solve the association problem. Note that we use both phase arrival time and amplitude information to model the Gaussian distributions.}
    \label{fig:GaMMA}
\end{figure}

We incorporate the physical constraints on phase arrival time and amplitude into a Gaussian mixture model to make it suitable for our association problem. \Cref{fig:GaMMA} illustrates how we model the Gaussian distributions to calculate the probability of the sequence of phases generated by the two causative earthquakes. The mathematical details of GaMMA are explained in the following sections.

\subsection{Bayesian Gaussian Mixture model}

We cast the association problem in a probabilistic framework where we use a Gaussian mixture distribution to model the probability of each phase pick:
\begin{align}
    p(\mathbf{x}_i) &= w_i \sum_{k=1}^K \phi_k \mathcal{N}(\mathbf{x}_i |\mathbf{\mu}_{k}, \mathbf{\Lambda}_k^{-1}) \\
    \mathcal{N}(\mathbf{x}_i | \mathbf{\mu}_{k}, \mathbf{\Lambda}_k^{-1}) &= \frac{1}{(2\pi)^{n/2}}|\mathbf{\Lambda}_k|^{1/2} \exp{\left( - \frac{1}{2} (\mathbf{x_i}-\mathbf{\mu}_{k})^T \mathbf{\Lambda}_k (\mathbf{x_i}-\mathbf{\mu}_{k}) \right)} \\
    \sum_{k=1}^K \phi_k &= 1
\end{align}
where $\mathbf{x}_i$ represents a phase pick including arrival time, phase type, and amplitude $(t_i, a_i)$ values at the $i$-the station. $\phi_k$ is the mixture component coefficient of the $k$-th earthquake. $\mathcal{N}$ represents a Gaussian distribution, and $\mathbf{\mu}_k$ is the mean of the Gaussian distribution. $\mathbf{\mu}_k$ represents the theoretical phase arrival time and amplitude ($\hat{t}_{ik}, \hat{a}_{ik}$) on each $i$-th station, determined by the $k$-th earthquake. $\mathbf{\Lambda}_k$ is the precision (inverse covariance) matrix of the Gaussian distribution. $w_i$ is the phase picking quality score between [0, 1]. $n$ is the number of feature dimensions, which is 1 if only time information is used or 2 if both time and amplitude information are used. Based on the Gaussian mixture distribution, we can calculate the probability of a set of recorded phases ($x_1, x_2, ..., x_N$). We assume these observations ($\mathbf{X}$) are independent and identically distributed (i.i.d.), then the log likelihood function is given by:
\begin{equation} \label{eq: log likelihood}
    \log\left(p(\mathbf{X}|\phi, \mathbf{\mu}, \mathbf{\Lambda})\right) = \sum_{i=1}^N \log \left( w_i \sum_{k=1}^K \phi_k \mathcal{N}(\mathbf{x}_i |\mathbf{\mu}_{k}, \mathbf{\Lambda}_k^{-1}) \right)
\end{equation}
We can find the assignment from $N$ phase picks to $K$ earthquakes, which is the goal of association, and the corresponding earthquake source parameters by maximizing the log likelihood of \eqref{eq: log likelihood}.

A limitation of the Gaussian mixture model formulation is that we need to assume the number of underlying earthquakes $K$. To address this unknown, we implement the Bayesian Gaussian mixture model \citep{bishop2006pattern}, which uses variational inference to calculate approximate posterior distributions for the parameters of a Gaussian mixture distribution.
Three conjugate priors are introduced. In particular, we use a Dirichlet prior for the mixture component coefficient, $p(\phi) = \mathcal{D}(\alpha_0)$ \citep{ferguson1973bayesian}, which controls the concentration of mixture components; a Gaussian prior for the mean conditioned on the precision, $p(\mu_k|\Lambda_k) = \mathcal{N}(m_0, \beta_0\Lambda_k)$; and a Wishart prior for the precision, $p(\Lambda_k) = \mathbf{W}(\textbf{W}_0, \nu_0)$ \citep{wishart1928generalised}, which controls the estimation of covariance.
The Bayesian model penalizes parameters that are away from the priors, which balances data fitting and model complexity. The mixture components (i.e., earthquakes) that do not contribute to the explaining the data (i.e., picks) will have approximately zero mixture coefficients so that we can choose a large number of components in the mixture model without over-fitting. In practice, we can initialize the space with many redundant earthquake hypocenters, and the Bayesian GMM suppresses unnecessary sources to infer a accurate number of earthquakes with associated picks.

\subsection{Expectation-Maximization (EM) algorithm} \label{sec:em}

We use the the Expectation-Maximization (EM) algorithm to solve the maximum likelihood estimation of $p(\mathbf{x})$. To consider the physical constraints on phase arrival time and amplitude for association, we incorporate the estimate of earthquake location, origin time, and magnitude into the EM algorithm. We then iteratively update the assignments from picks to earthquakes in the E-step and optimize the earthquake parameters in the M-step:
\paragraph{E-step:}
\begin{equation}
    \gamma_{ik} = \frac{\phi_k \mathcal{N}(\mathbf{x}_i | \mathbf{\mu}_k, \mathbf{\Sigma}_k)}{\sum_{k=1}^K \phi_k \mathcal{N}(\mathbf{x}_i | \mathbf{\mu}_k, \mathbf{\Sigma}_k)} 
    \label{eqn:e_step}
\end{equation}
where $\gamma_{ik}$ is the probability that phase pick $\mathbf{x}_i$, i.e., arrival time $t_{i}$ and wave amplitude $a_{i}$, is generated by the $k$-th earthquake.
\paragraph{M-step:}
\begin{enumerate}
    \item Effective number of picks assigned to the $k$-th earthquake:
\begin{align}
    N_k &= \sum_{i=1}^N \gamma_{ik} \\
    \phi_k &= \frac{N_k}{N}
\end{align}
    \item Earthquake location, origin time, and magnitude of the $k$-th earthquake:
\begin{align}
    \minimize_{(x_k, y_k, z_k, t_k)} l(x_k, y_k, z_k, t_k) &= \sum_{i=1}^N  \gamma_{ik} \mathcal{L}\left( t_{i}, \hat{t}_{ik}(x_k, y_k, z_k, t_k) \right) \label{eqn:eq_loc} \\
    m_k &= \frac{1}{N_k} \sum_{i=1}^N \gamma_{ik} \mathcal{F}'_a(a_i, d_{ik}) \label{eqn:eq_mag}
\end{align}
    \item Theoretical travel time, amplitude, and  statistics of residuals:
\begin{align}
    \mathbf{\mu}_{k} &= \left[\begin{array}{c} \hat{t}_{ik} \\ \hat{a}_{ik}\end{array}\right] = \left[\begin{array}{c} \mathcal{F}_t(x_k, y_k, z_k, t_k) \\ \mathcal{F}_a(m_k, d_{ik})\end{array}\right]\\
    \mathbf{\Lambda}^{-1}_k &= \frac{1}{N_k} \sum_{i=1}^N \gamma_{ik} (\mathbf{x}_i - \mathbf{\mu}_k)(\mathbf{x}_i - \mathbf{\mu}_k)^T \label{eqn:sigma}
\end{align}
\end{enumerate}
where $\mathcal{L}$ is a loss function of the residuals between the picked phase arrival time $t_i$ and the theoretical arrival time $\hat{t}_{ik}$ from the $k$-th earthquake. Minimization of the loss function $l$ gives an estimate of the earthquake location and origin time $(x_k, y_k, z_k, t_k)$. $m_k$ is the magnitude of the $k$-th earthquake. $\mathcal{F}_t$ represents the function used to calculate theoretical phase arrival time $\hat{t}_{ik}$. $\mathcal{F}_a$ represents the function to calculate theoretical phase amplitude $\hat{a}_{ik}$ based on earthquake magnitude $m_k$, and $\mathcal{F}'_a$ represents the function to estimate earthquake magnitude using phase amplitude. $d_{ik}$ is the distance from the $k$-th earthquake to $i$-th seismic station.
Here we decouple the optimization of earthquake magnitude (\Cref{eqn:eq_mag}) from the optimization of earthquake location and time (\Cref{eqn:eq_loc}). 
We use arrival times to constrain earthquake location and use phase amplitudes to constrain earthquake magnitude. 
Note that although we decouple the two optimizations, the precision matrix (\Cref{eqn:sigma}) considers the correlation between arrival time and amplitude residuals. In this way, both arrival time and amplitude information are used for the association process of clustering picks among earthquakes (\Cref{eqn:e_step}).

For the Bayesian Gaussian mixture model, we add another stage in the M-step to update the posterior parameters:
\begin{align}
    \alpha_{k} &=\alpha_{0}+N_{k} \\
    \beta_{k} &=\beta_{0}+N_{k} \\
    \mathbf{m}_{k} &=\frac{1}{\beta_{k}}\left(\beta_{0} \mathbf{m}_{0}+N_{k} \mathbf{\mu}_{k}\right) \\
    \mathbf{W}_{k}^{-1} &=\mathbf{W}_{0}^{-1}+N_{k} \mathbf{\Lambda}^{-1}_k+\frac{\beta_{0} N_{k}}{\beta_{0}+N_{k}}\left(\mathbf{\mu}_{k}-\mathbf{m}_{0}\right)\left(\mathbf{\mu}_{k}-\mathbf{m}_{0}\right)^{\mathrm{T}} \\
    \nu_{k} &=\nu_{0}+N_{k}
\end{align}
The E-step is modified as:
\begin{align}
    \gamma_{i k} &\propto \widetilde{\pi}_{k} \widetilde{\Lambda}_{k}^{1 / 2} \exp \left\{-\frac{D}{2 \beta_{k}}-\frac{\nu_{k}}{2}\left(\mathbf{x}_{n}-\mathbf{m}_{k}\right)^{\mathrm{T}} \mathbf{W}_{k}\left(\mathbf{x}_{n}-\mathbf{m}_{k}\right)\right\} \\
    \ln \widetilde{\Lambda}_{k} &=\sum_{i=1}^{D} \psi\left(\frac{\nu_{k}+1-i}{2}\right)+D \ln 2+\ln \left|\mathbf{W}_{k}\right| \\
    \ln \widetilde{\pi}_{k} &=\psi\left(\alpha_{k}\right)-\psi(\widehat{\alpha})
\end{align}
where $\widehat{\alpha} = \sum_k \alpha_k$ and $\psi$ is the digamma function \citep{abramowitz1964handbook}.
The explanation of these updating rules is detailed in  \citet{bishop2006pattern}'s textbook.

\subsection{Earthquake location and magnitude estimation}

The iteration of the EM algorithm both updates the clustering of picks based on earthquakes and optimizes the corresponding earthquake source parameters. We focus on efficient association rather than accuracy of earthquake source parameters, which can be realized once phases are properly associated, so we choose two basic approaches to estimate approximate earthquake locations and magnitudes. We optimize \Cref{eqn:eq_loc} with a Huber loss function \citep{huber1992robust} as the target to reduce the effect of outliers:
\begin{equation}
\mathcal{L}_{\delta}(t - \hat{t})=\left\{\begin{array}{ll}
\frac{1}{2} (t - \hat{t})^{2} & \text { for }|t - \hat{t}| \leq \delta \\
\delta\left(|t - \hat{t}|-\frac{1}{2} \delta\right), & \text { otherwise. }
\end{array}\right.
\end{equation}
where the hyper-parameter $\delta$ is set to 1 second in this test. For this proof-of-concept study, we use a uniform velocity model to calculate the theoretical phase travel-time:
\begin{equation}
    \widehat{t}_{ik}(x_k, y_k, z_k, t_k) = \mathcal{F}_t(x_k, y_k, z_k, t_k) = \frac{d_{ik}}{v} + t_k
\end{equation}
We then solve the minimization of \Cref{eqn:eq_loc} using the BFGS algorithm \citep{fletcher2013practical}. 
Advanced earthquake location algorithms and complex velocity models can also be applied to solving \Cref{eqn:eq_loc} but at a higher computational cost.

To estimate earthquake magnitude in \Cref{eqn:eq_mag}, we use a linear relationship between log phase amplitude and earthquake magnitude:
\begin{equation}
    \widehat{m}_{ik} = \mathcal{F}'_a(a_i, d_{ik}) = c_0 + c_1 \log a_i + c_2 \log d_{ik}
    \label{eqn:gmpe}
\end{equation}
Station correction terms can been added to consider site effects, i.e.  site amplification factor \citep{munchmeyer2020low}. Based on the measured phase amplitude type, e.g., displacement, peak ground velocity, or peak ground acceleration, we can choose from among the Richter empirical magnitude relationship \citep{richter1935instrumental}, the Richter simulation-based prediction \citep{AlIsmael2020}, or a simplified ground motion prediction equations \citep{picozzi2018rapid} for \Cref{eqn:gmpe}.

\section{Results}

We demonstrate the performance of GaMMA first on a synthetic example and then on six days of data from the 2019 Ridgecrest, California earthquake sequence.

\subsection{Synthetic test}

\begin{figure}
    \centering
    \begin{subfigure}{\textwidth}
    \includegraphics[width=\textwidth]{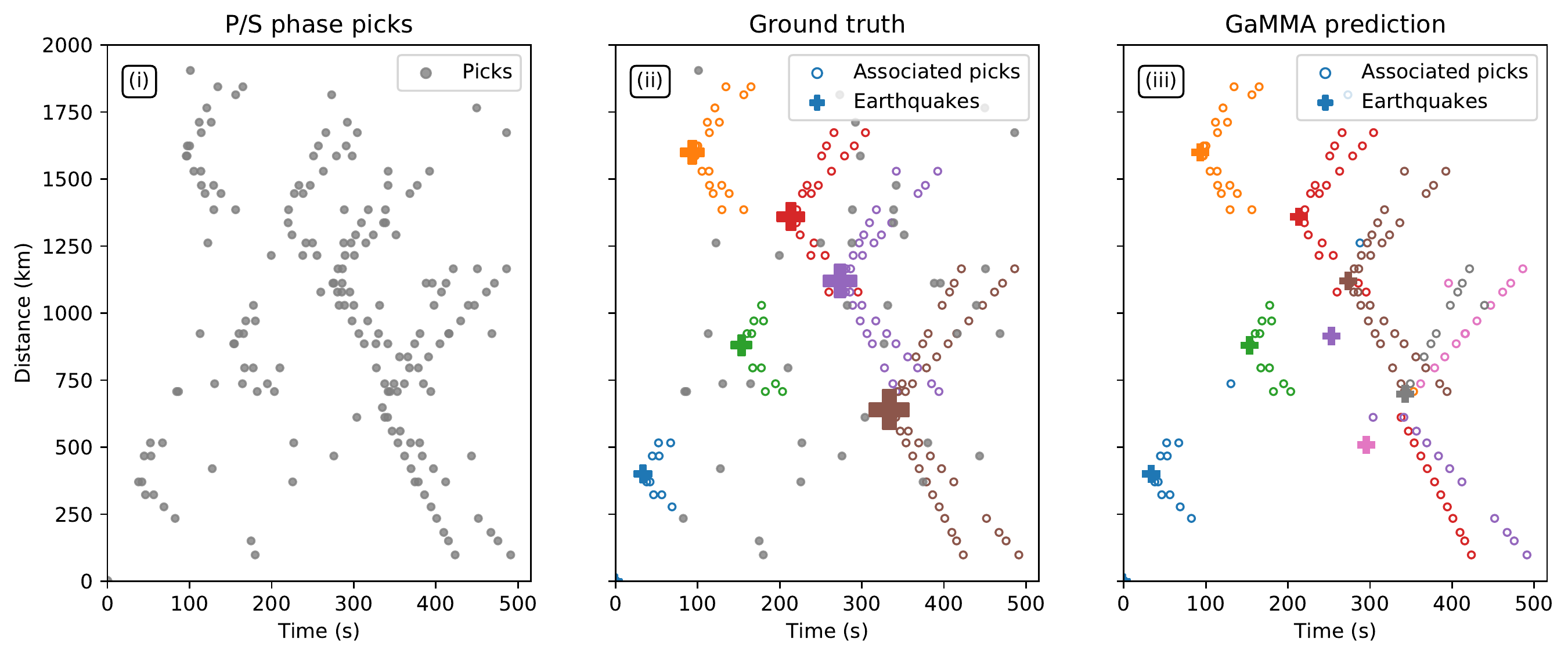}
    \caption{}
    \end{subfigure}
    \begin{subfigure}{\textwidth}
    \includegraphics[width=\textwidth]{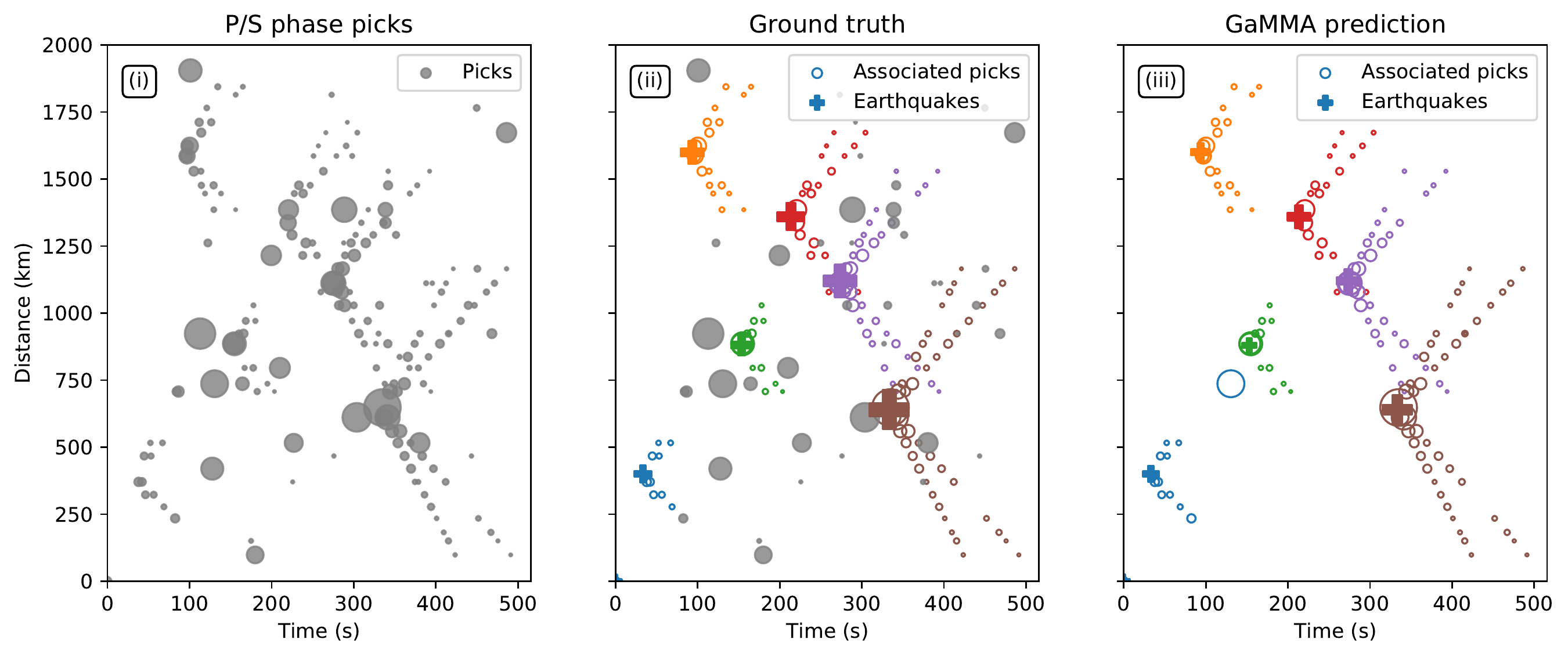}
    \caption{}
    \end{subfigure}
    \caption{Synthetic example: (a) association using only time; (b) association using both time and amplitude. The left panels plot the P- and S-phase picks. The middle panels are plots of the ground truth of association result. The unassociated false positives are plotted in grey. The circle size represents phase amplitude and the cross size represents earthquake magnitude. The right panels show the association results of the GaMMA method. Note that some phases in the lower right corner of panel (a)(iii) are mis-associated with another distant earthquake marked in red, because these phases can fit the moveout of both events. Amplitude information provides an extra constraint in distance that resolves this ambiguity as shown in panel (b)(iii).}
    \label{fig:synthetic}
\end{figure}

We first created a synthetic experiment to demonstrate the association results of GaMMA. We generated a sequence of phases including both P- and S- phases from six earthquake events. To model the errors that exist in real data, we added a 0.5s random error in the phase arrival times and scaled the phase amplitude (peak ground velocity (PGV)) by a random factor between 0.3-3. We further added 30\% false positive picks at random times. In total, 178 P- and S-phase picks from 40 stations were used for association (left panels of \Cref{fig:synthetic}). We used a simplified ground motion prediction equation of PGV from \citet{picozzi2018rapid}'s work: $\log PGV = -2.175 - 1.68 \log R + 0.93 M$, where R is hypocentral distance and M is earthquake magnitude. The ground truth result is shown in the middle panels of \Cref{fig:synthetic}. The symbol size represents the relative size of phase amplitude and earthquake magnitude.
We conducted two association tests using in the first test only the arrival time information (\Cref{fig:synthetic}(a)) and in the second test both the arrival time and amplitude information (\Cref{fig:synthetic}(b)). 
The same parameters and initialization were used for both cases. For each we initialized the earthquake locations at the center of research area and uniformly distributed the earthquake origin times. The association results are shown in the right panels of \Cref{fig:synthetic}. At least five of the six true earthquakes are successfully associated in both tests. However, the sixth event in the lower right corner of \Cref{fig:synthetic}(a, b)(ii) (marked in brown) is successfully associated only when amplitudes are used in conjunction with travel times (\Cref{fig:synthetic}(b)(iii)). Without amplitudes (\Cref{fig:synthetic}(a)(iii)), the result includes multiple incorrect event associations marked by colors pink and purple. This occurs because the P-phases of this event have arrival times that overlap with another distant event's moveout (the event marked in red) and could be mistakenly associated with this distant event. The extra information provided by amplitude adds the necessary extra constraint on distance, in addition to time, that allows the sixth event to be associated correctly. The test with amplitude information also correctly estimates the earthquake magnitudes. This synthetic experiment demonstrates that GaMMA benefits from using both the time and amplitude information in the association process.

\subsection{Test on the 2019 Ridgecrest earthquake sequence}

We next applied the GaMMA method to part of the 2019 Ridgecrest earthquake sequence to evaluate its performance on a real earthquake sequence. We focused on the initial six days of the sequence when a large number of earthquakes occurred and migrated from the southwest-striking fault to the northwest-striking fault. 
We applied the PhaseNet model \citep{zhu2019phasenet} to extract picks from waveforms of ``HH," ``BH," ``EH," and ``HN" channels of 23 stations of the ``CI" network within 1 degree of the location (-117.504$W$, 35.705$N$). We measured the PGV value over a 8s window after the phase arrival time. We then associated the detected 651,994 P-picks and 686,291 S-picks using the GaMMA method. We used a uniform velocity model with $v_p = 6$ km/s and $v_s = v_p/1.75$ for earthquake location estimation and a simple ground motion prediction equation as above for earthquake magnitude estimation. Because we used such a simple moveout behavior, we do not expect the earthquake locations to be highly accurate, but they are close enough for successful association, given the relatively short source-receiver distances involved.

\begin{figure}
    \centering
    \begin{subfigure}{0.48\textwidth}
    \includegraphics[width=\textwidth]{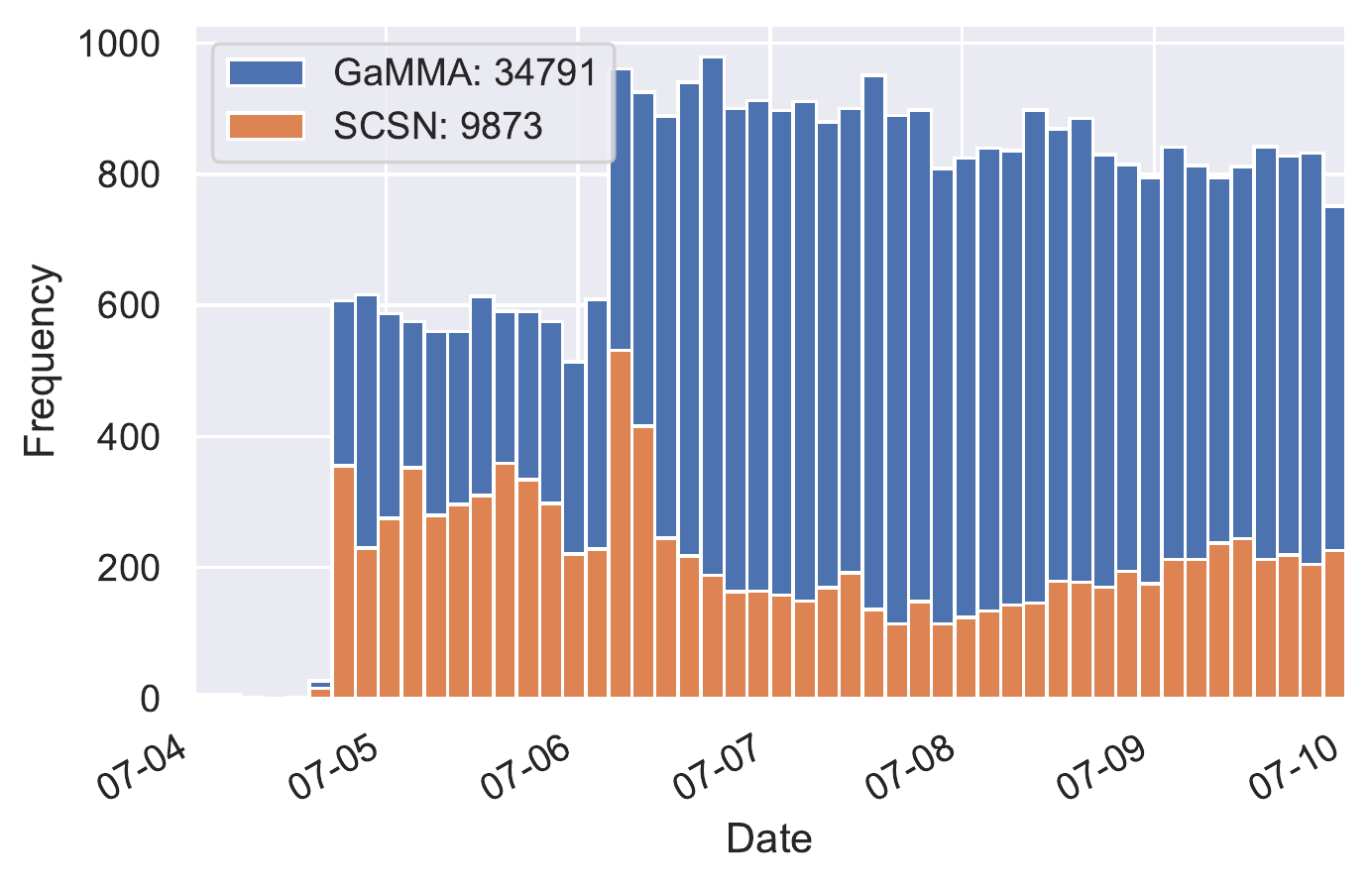}
    \caption{}
    \end{subfigure}
    \begin{subfigure}{0.48\textwidth}
    \includegraphics[width=\textwidth]{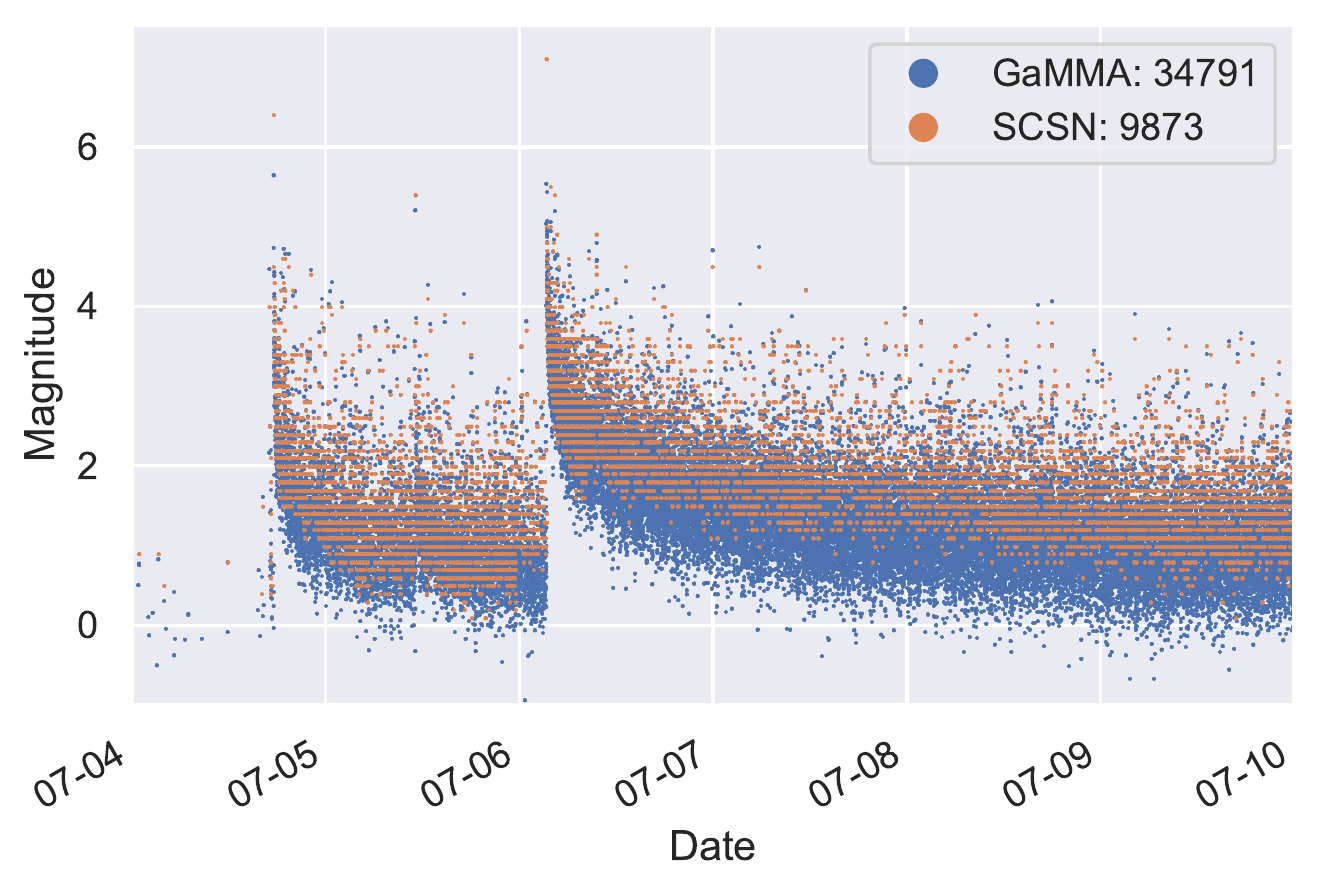}
    \caption{}
    \end{subfigure}
    \begin{subfigure}{1.0\textwidth}
    \includegraphics[width=\textwidth]{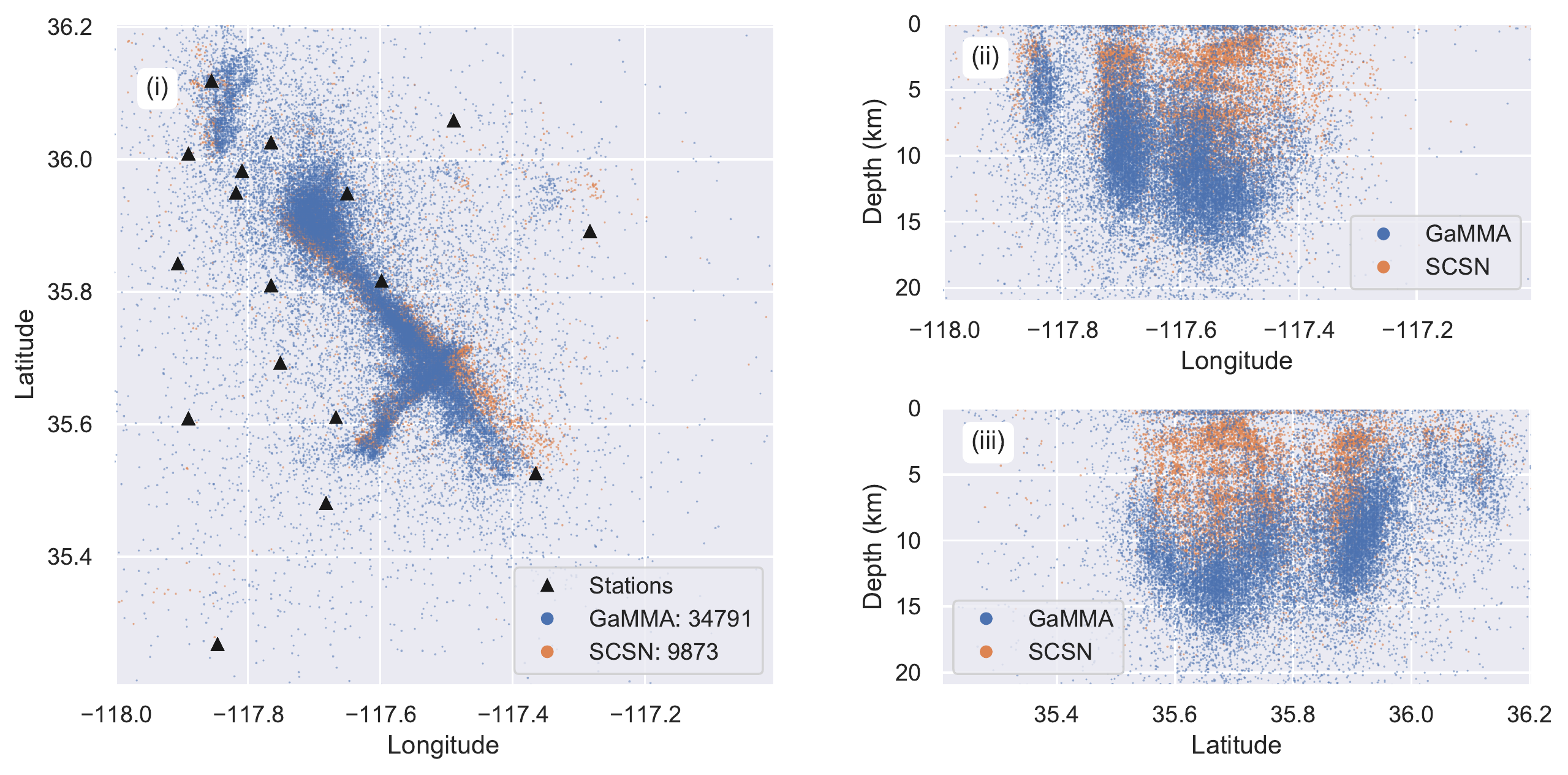}
    \caption{}
    \end{subfigure}
    \caption{Association results of the Ridgecrest dataset: (a) associated earthquake frequency, (b) associated earthquake magnitude, (c) associated earthquake location. Note that because we use a uniform velocity model during association, we do not expect the earthquake locations to be accurate, but they are close enough for effective association.}
    \label{fig:ridgecrest}
\end{figure}

\begin{figure}
    \centering
    \includegraphics[width=1.0\textwidth]{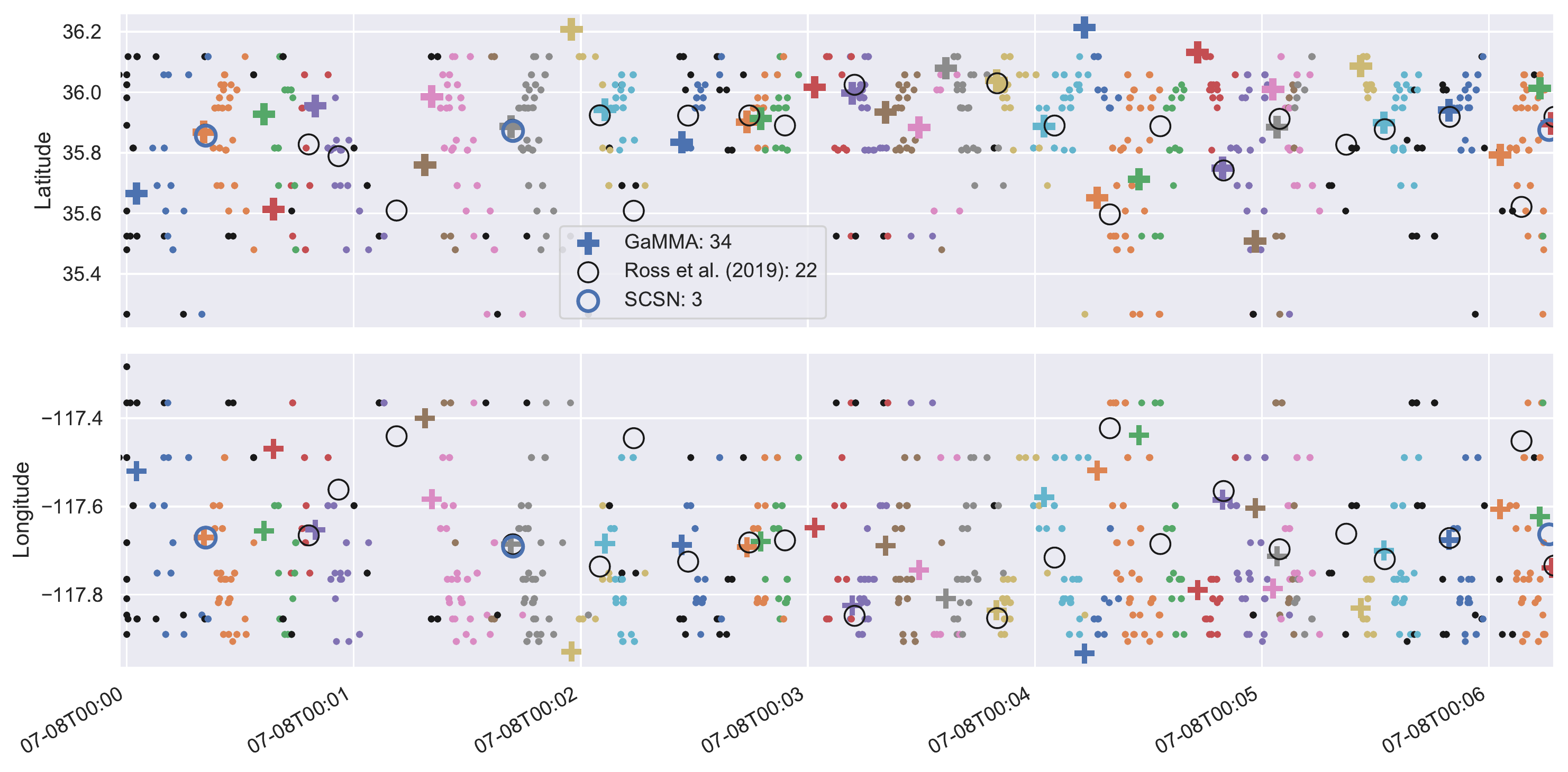}
    \caption{An example of association results from a dense sequence of phase picks starting at time 2019-07-08T00:00:00 (UTC). GaMMA associates 32 events during this period, while there are only 3 events in the SCSN catalog and 22 events in Ross et al. (2019)'s template matching catalog.}
    \label{fig:example}
\end{figure}

\begin{figure}
    \centering
    \includegraphics[width=1.0\textwidth]{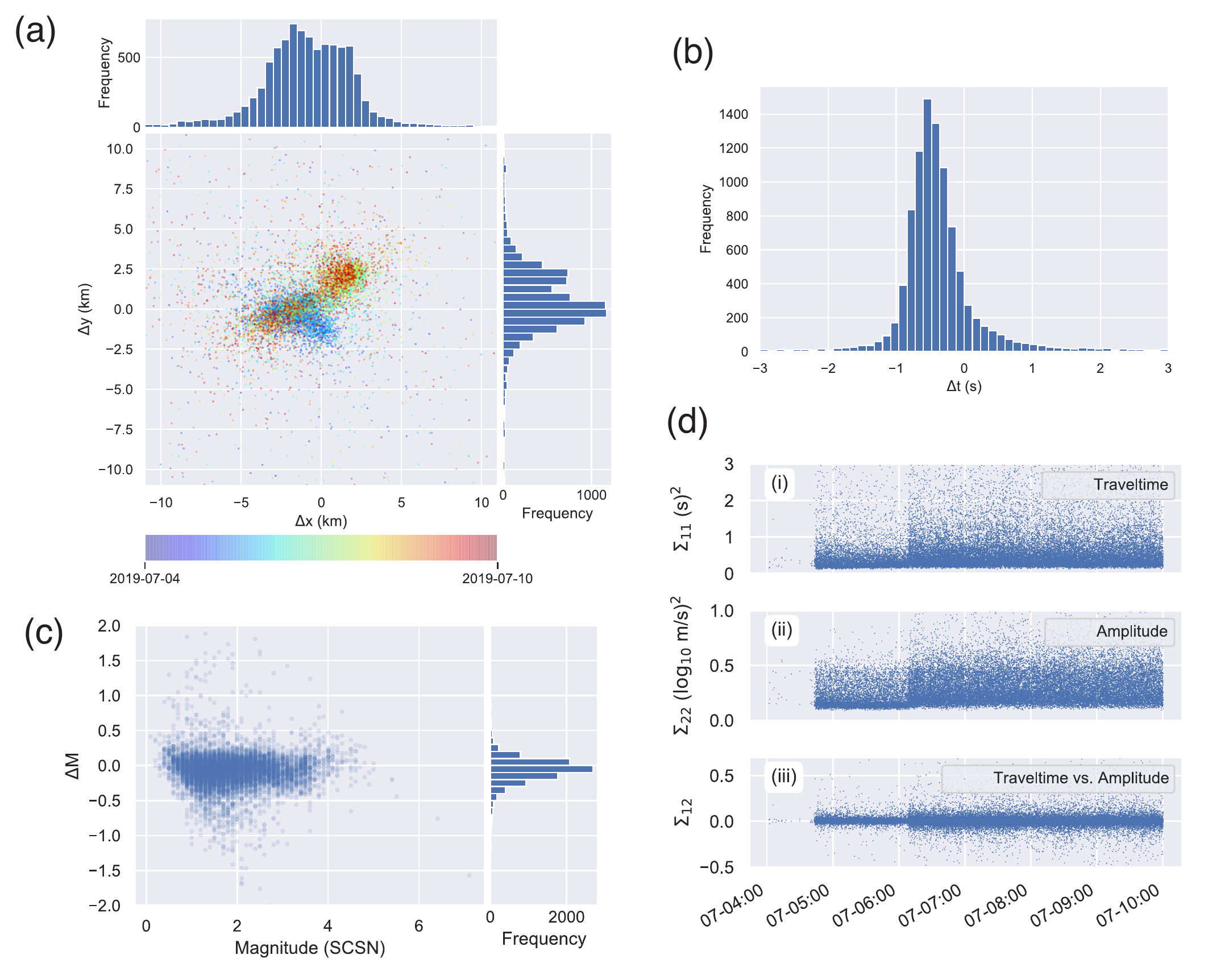}
    \caption{Residuals of (a) associated earthquake location, (b) origin time, and (c) magnitude compared with the SCSN catalog. (d) The components of the covariance matrix of travel-time and amplitude residuals estimated by GaMMA.}
    \label{fig:error}
\end{figure}

\Cref{fig:ridgecrest} shows the statistics of the 34,791 associated earthquakes from 598,218 P-picks and 633,010 S-picks, leaving 53,776 P-picks and 53,281 S-picks unassociated. 
We also plotted the 9,873 earthquakes in the SCSN catalog \citep{southern2013southern} for comparison. 
Both the associated earthquake locations and magnitudes agree with the SCSN catalog (\Cref{fig:ridgecrest}(b) and (c)). 
\Cref{fig:example} shows an association example with a dense sequence of picks occurring during a 8-minute period. GaMMA associates 32 events during this period, while there are only 3 events in the SCSN catalog and 22 events in \cite{ross2019hierarchical}'s template matching catalog.
\Cref{fig:error} shows the residual distributions of the associated earthquake location and magnitude compared with the SCSN catalog. The statistics of mean, standard deviation (STD), and median absolute error (MAE) can be found in \Cref{tab:error}. The covariance matrix in \Cref{fig:error}(d) shows that most of the associated earthquakes have small residuals of phase arrival time and amplitude, indicating that the phase picks match well with the theoretical values determined by the causative earthquakes found by association. 
These location and magnitude estimates can be further improved through the application of established earthquake location and magnitude algorithms once the picks have been associated.

\begin{figure}
    \centering
    \includegraphics[width=1.0\textwidth]{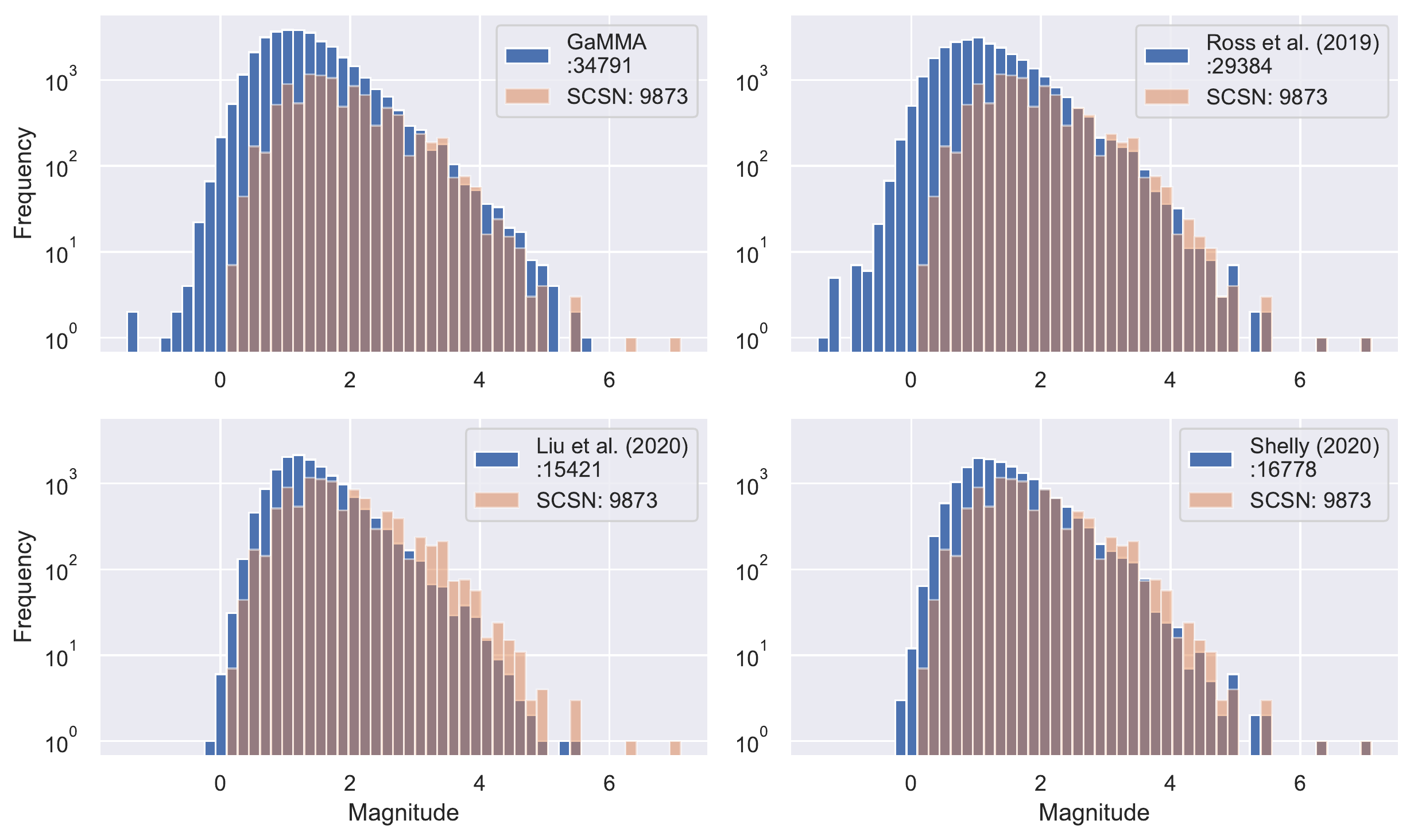}
    \caption{Comparison of magnitude distribution.}
    \label{fig:compare_mag}
\end{figure}

\begin{figure}
    \centering
    \includegraphics[width=1.0\textwidth]{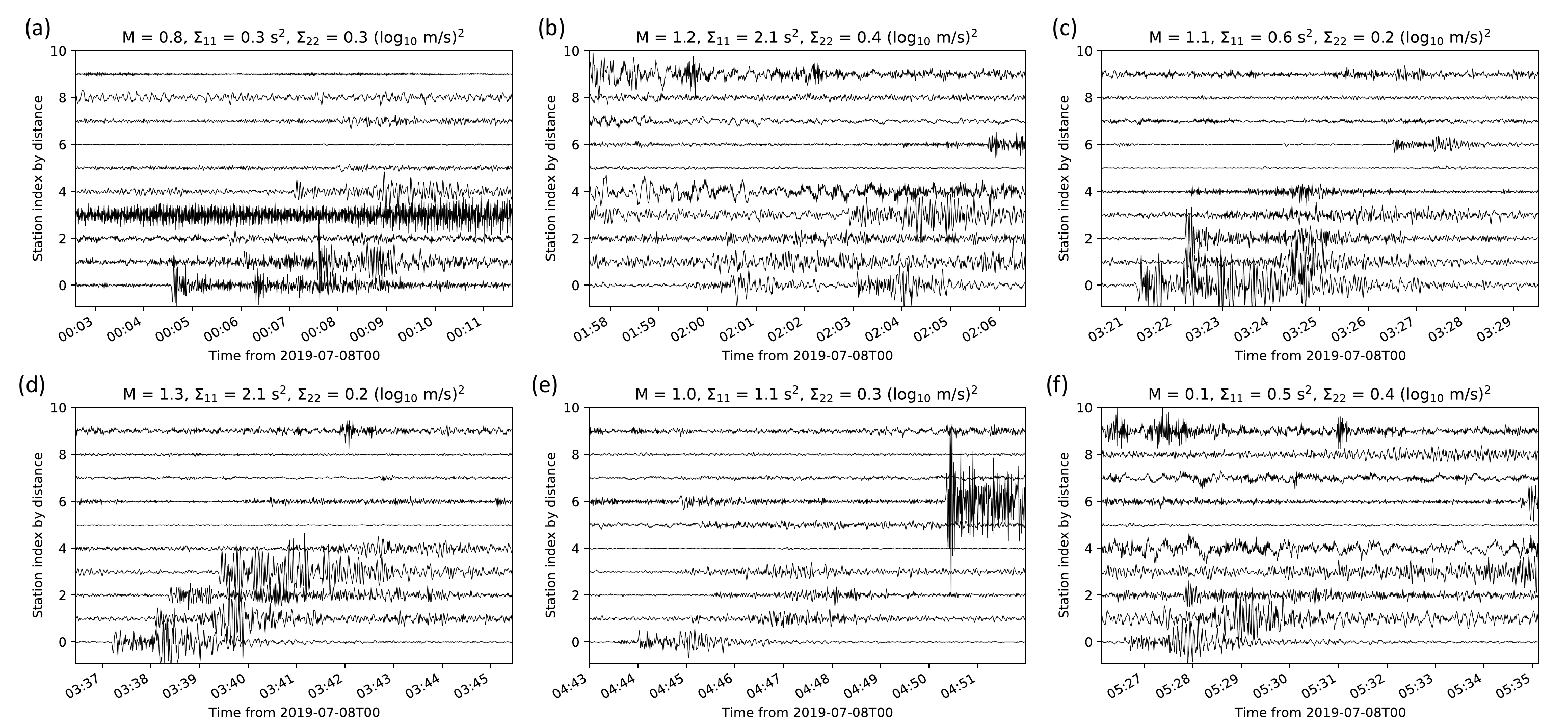}
    \caption{Waveforms of six newly detected events that are not Ross et al. (2019)'s template matching catalog in \Cref{fig:example}. M is the associated earthquake magnitude. $\Sigma_{11}$ and $\Sigma_{22}$ are the variances of associated phase time and phase amplitude respectively. These earthquakes have a small magnitude and can only be detected at a few stations. We can also see a missed earthquake event at around 2019-07-08T00:02:03 in (b).}
    \label{fig:waveform}
\end{figure}

We compared the catalog generated by GaMMA with three state-of-art catalogs \citep{ross2019hierarchical, liu2020rapid, shelly2020high}. \Cref{tab:compare_others} shows the earthquake numbers in these catalogs during the same period. In each of these cases we assumed these catalogs as ground truth and analyzed whether the earthquakes they contain are also detected in GaMMA's catalog within a 5s window. Based on the recall rate, more than 95\% of earthquakes in the catalogs of SCSN, \citet{liu2020rapid}, and \citet{shelly2020high} are successfully associated by GaMMA. The low precision and F1-score are due to the large number of new earthquakes associated by GaMMA. To verity whether these new earthquakes are reasonable, we compared the magnitude distributions of the four catalogs (\Cref{fig:compare_mag}). Most of these new earthquakes associated by GaMMA have a small magnitude and follow the Gutenberg–Richter magnitude-frequency relationship \citep{gutenberg1956energy}, which suggests that they may be legitimate detections of real earthquakes. \Cref{fig:waveform} shows seismic waveforms of six newly detected events in \Cref{fig:example}. We can see clear earthquake signals in these examples. Meanwhile, these signals are relatively weak and can only be detected at a few stations. Comprehensive comparison among these catalogs is a subject of future research.

\begin{table}[]
\centering
\caption{Statistics of residual distributions in \Cref{fig:error}}
\label{tab:error}
\resizebox{0.6\textwidth}{!}{%
\begin{tabular}{cccccc}
\hline
Error & $\Delta x$ (km) & $\Delta y$ (km) & $\Delta z$ (km) & $\Delta t$ (s) & $\Delta m$ \\ \hline
Mean  & -0.85           & 0.45            & 10.75           & -0.35          & -0.064     \\
STD   & 2.61            & 2.11            & 3.80            & 0.80           & 0.235      \\
MAE   & 2.13            & 1.53            & 10.75           & 0.62           & 0.154      \\ \hline
\end{tabular}%
}
\end{table}

\begin{table}[]
\centering
\caption{Comparison with other catalogs}
\label{tab:compare_others}
\resizebox{0.8\textwidth}{!}{%
\begin{tabular}{cccccc}
\hline
\multicolumn{2}{c}{Catalog} &
  \begin{tabular}[c]{@{}c@{}}SCSN\\ (9,873)\end{tabular} &
  \begin{tabular}[c]{@{}c@{}}Ross et al. (2019)\\ (29,384)\end{tabular} &
  \begin{tabular}[c]{@{}c@{}}Liu et al. (2020)\\ (15,421)\end{tabular} &
  \begin{tabular}[c]{@{}c@{}}Shelly (2020)\\ (16,778)\end{tabular} \\ \hline
\multirow{3}{*}{\begin{tabular}[c]{@{}c@{}}GaMMA\\ (34,791)\end{tabular}} & Recall    & 0.973 & 0.737 & 0.987 & 0.955 \\ \cline{2-2}
                                                                          & Precision & 0.336 & 0.765 & 0.576 & 0.552 \\ \cline{2-2}
                                                                          & F1-score  & 0.499 & 0.751 & 0.727 & 0.699 \\ \hline
\end{tabular}%
}
\end{table}

\section{Discussion}

The Gaussian mixture model (GMM) is an effective and widely used unsupervised learning method for clustering. We have combined the GMM with earthquake location and magnitude estimation to develop a novel seismic phase association method (GaMMA). We treat earthquake phase association as a clustering problem that aims to cluster phases based on the causative earthquakes. There are several advantageous features of this unsupervised learning approach. First, GaMMA does not require grid-search or training commonly used in other association methods. Second, GaMMA can flexibly consider phase arrival time, amplitude, P/S type, and pick quality. While it is difficult for conventional association methods to consider phase amplitude information, GaMMA can easily include phase amplitude information both to improve association and to estimate earthquake magnitude. Finally, GaMMA optimizes the association result in a probabilistic framework and estimates the covariance of time and amplitude residuals. These advantages suggest that GaMMA is a promising approach for improved earthquake phase association, which is in turn an important component of improved earthquake monitoring. 

We note that there are several limitations of GaMMA that need to be considered. 
First, the time complexity of the Gaussian mixture model scales with $O(K \cdot N)$, where $K$ is the number of clusters (earthquakes) and $N$ is the number of samples (phase picks), so the computational cost could become prohibitive for a long earthquake sequence. In practice, however, it is straightforward and effective to segment a long sequence into relatively shorter windows to improve the association speed by processing data in parallel, as phases that are separated by a certain time interval (depending on the source-receiver distance) cannot come from a single earthquake. In this work, we used the DBSCAN algorithm \citep{schubert2017dbscan} to divide picks into sub-windows for association. The example in \Cref{fig:example} shows one sub-window from the seven hour sequence.
Second, the clustering results of the Gaussian mixture model are affected by the initialization state. In this work, we used a simple strategy to initialize the earthquake locations uniformly in time and space. The number of initialized earthquakes is proportional to the ratio between the number of phase picks divided and the number of stations. This simple strategy worked well in the experiments described above. Improving initialization strategies has the potential to increase the association performance further.
Third, GaMMA ensures that one pick is only assigned to one earthquake, while conventional back-projection methods may attribute one pick to several earthquakes; however, GaMMA does not consider station-based constraints, such as that only one pair of P- or S-picks from one station is assigned to each earthquake. \citet{mcbrearty2019earthquake} accounts for this constraint using a constrained ILP (integer linear programming) solution in their association method, and there is a strong correspondence between that technique and our technique. One potential solution that would allow introduction of the station-based constraint is to add a normalizing scheme similar to \Cref{eqn:e_step} over stations. 
Finally, GaMMA assumes that the residuals of pick time or amplitude follow a Gaussian distribution. This assumption is not accurate for false positive picks. Adding a mixture component of background uniform distribution to account false positive picks might be a helpful extension to be considered in future research \citep{melchior2018filling}.

\section{Conclusions}
We have developed a new association method based on a Bayesian Gaussian mixture model, GaMMA, which solves the phase association problem as an unsupervised clustering problem. To consider the physical constraints of phase arrival time and amplitude with earthquake location and magnitude, we incorporate optimization of earthquake location and magnitude into the Expectation-Maximization (EM) algorithm commonly used for solving the Gaussian mixture model. GaMMA, thus, can use both arrival time and amplitude information to cluster picks from the same earthquake and simultaneously estimates both the earthquake location and magnitude from each cluster of picks. The experiment results on both synthetic tests and the 2019 Ridgecrest earthquake sequence demonstrate the effectiveness of the GaMMA method in associating a dense sequence of P- and S-phase picks. GaMMA provides an unsupervised learning approach to solve the challenging phase association problem resulting from the increasingly wide applications of deep-learning-based phase pickers. The improved performance of GaMMA can associate more earthquakes from massive automatic phase picks, thus enriching earthquake catalogs and improving earthquake monitoring.

\section*{Acknowledgements}
The phase picking and association data are available in Open Science Framework (\url{https://doi.org/10.17605/OSF.IO/3GP72}). The code is open source in GitHub (\url{https://github.com/wayneweiqiang/GMMA}). GaMMA is developed based on the scikit-learn package (\url{https://github.com/scikit-learn/scikit-learn}).
The data of 2019 Ridgecrest earthquake can be accessed from Southern California Earthquake Data Center.
This work is supported by the Department of Energy Basic Energy Sciences (DE-SC0020445).

\clearpage
\bibliographystyle{apacite}
\bibliography{reference}

\begin{thebibliography}{}

\bibitem [\protect \citeauthoryear {%
Abramowitz%
\ \BBA {} Stegun%
}{%
Abramowitz%
\ \BBA {} Stegun%
}{%
{\protect \APACyear {1964}}%
}]{%
abramowitz1964handbook}
\APACinsertmetastar {%
abramowitz1964handbook}%
\begin{APACrefauthors}%
Abramowitz, M.%
\BCBT {}\ \BBA {} Stegun, I\BPBI A.%
\end{APACrefauthors}%
\unskip\
\newblock
\APACrefYear{1964}.
\newblock
\APACrefbtitle {Handbook of mathematical functions with formulas, graphs, and
  mathematical tables} {Handbook of mathematical functions with formulas,
  graphs, and mathematical tables}\ (\BVOL~55).
\newblock
\APACaddressPublisher{}{US Government printing office}.
\PrintBackRefs{\CurrentBib}

\bibitem [\protect \citeauthoryear {%
Allen%
}{%
Allen%
}{%
{\protect \APACyear {1978}}%
}]{%
allen1978automatic}
\APACinsertmetastar {%
allen1978automatic}%
\begin{APACrefauthors}%
Allen, R\BPBI V.%
\end{APACrefauthors}%
\unskip\
\newblock
\APACrefYearMonthDay{1978}{}{}.
\newblock
{\BBOQ}\APACrefatitle {Automatic earthquake recognition and timing from single
  traces} {Automatic earthquake recognition and timing from single
  traces}.{\BBCQ}
\newblock
\APACjournalVolNumPages{Bulletin of the Seismological Society of
  America}{68}{5}{1521--1532}.
\PrintBackRefs{\CurrentBib}

\bibitem [\protect \citeauthoryear {%
Al‐Ismail%
, Ellsworth%
\BCBL {}\ \BBA {} Beroza%
}{%
Al‐Ismail%
\ \protect \BOthers {.}}{%
{\protect \APACyear {2020}}%
}]{%
AlIsmael2020}
\APACinsertmetastar {%
AlIsmael2020}%
\begin{APACrefauthors}%
Al‐Ismail, F.%
, Ellsworth, W\BPBI L.%
\BCBL {}\ \BBA {} Beroza, G\BPBI C.%
\end{APACrefauthors}%
\unskip\
\newblock
\APACrefYearMonthDay{2020}{03}{}.
\newblock
{\BBOQ}\APACrefatitle {{Empirical and Synthetic Approaches to the Calibration
  of the Local Magnitude Scale, ML, in Southern Kansas}} {{Empirical and
  Synthetic Approaches to the Calibration of the Local Magnitude Scale, ML, in
  Southern Kansas}}.{\BBCQ}
\newblock
\APACjournalVolNumPages{Bulletin of the Seismological Society of
  America}{110}{2}{689-697}.
\PrintBackRefs{\CurrentBib}

\bibitem [\protect \citeauthoryear {%
Barrett%
\ \BBA {} Beroza%
}{%
Barrett%
\ \BBA {} Beroza%
}{%
{\protect \APACyear {2014}}%
}]{%
barrett2014}
\APACinsertmetastar {%
barrett2014}%
\begin{APACrefauthors}%
Barrett, S.%
\BCBT {}\ \BBA {} Beroza, G.%
\end{APACrefauthors}%
\unskip\
\newblock
\APACrefYearMonthDay{2014}{05}{}.
\newblock
{\BBOQ}\APACrefatitle {An Empirical Approach to Subspace Detection} {An
  empirical approach to subspace detection}.{\BBCQ}
\newblock
\APACjournalVolNumPages{Seismological Research Letters}{85}{}{594-600}.
\newblock
\begin{APACrefDOI} \doi{10.1785/0220130152} \end{APACrefDOI}
\PrintBackRefs{\CurrentBib}

\bibitem [\protect \citeauthoryear {%
Beroza%
, Segou%
\BCBL {}\ \BBA {} Mostafa~Mousavi%
}{%
Beroza%
\ \protect \BOthers {.}}{%
{\protect \APACyear {2021}}%
}]{%
beroza2021machine}
\APACinsertmetastar {%
beroza2021machine}%
\begin{APACrefauthors}%
Beroza, G\BPBI C.%
, Segou, M.%
\BCBL {}\ \BBA {} Mostafa~Mousavi, S.%
\end{APACrefauthors}%
\unskip\
\newblock
\APACrefYearMonthDay{2021}{}{}.
\newblock
{\BBOQ}\APACrefatitle {Machine learning and earthquake forecasting—next
  steps} {Machine learning and earthquake forecasting—next steps}.{\BBCQ}
\newblock
\APACjournalVolNumPages{Nature Communications}{12}{1}{1--3}.
\PrintBackRefs{\CurrentBib}

\bibitem [\protect \citeauthoryear {%
Bishop%
}{%
Bishop%
}{%
{\protect \APACyear {2006}}%
}]{%
bishop2006pattern}
\APACinsertmetastar {%
bishop2006pattern}%
\begin{APACrefauthors}%
Bishop, C\BPBI M.%
\end{APACrefauthors}%
\unskip\
\newblock
\APACrefYear{2006}.
\newblock
\APACrefbtitle {Pattern Recognition and Machine Learning} {Pattern recognition
  and machine learning}.
\newblock
\APACaddressPublisher{{New York}}{{Springer}}.
\PrintBackRefs{\CurrentBib}

\bibitem [\protect \citeauthoryear {%
Dickey%
, Borghetti%
, Junek%
\BCBL {}\ \BBA {} Martin%
}{%
Dickey%
\ \protect \BOthers {.}}{%
{\protect \APACyear {2020}}%
}]{%
dickey2020beyond}
\APACinsertmetastar {%
dickey2020beyond}%
\begin{APACrefauthors}%
Dickey, J.%
, Borghetti, B.%
, Junek, W.%
\BCBL {}\ \BBA {} Martin, R.%
\end{APACrefauthors}%
\unskip\
\newblock
\APACrefYearMonthDay{2020}{}{}.
\newblock
{\BBOQ}\APACrefatitle {Beyond correlation: A path-invariant measure for
  seismogram similarity} {Beyond correlation: A path-invariant measure for
  seismogram similarity}.{\BBCQ}
\newblock
\APACjournalVolNumPages{Seismological Research Letters}{91}{1}{356--369}.
\PrintBackRefs{\CurrentBib}

\bibitem [\protect \citeauthoryear {%
Draelos%
, Ballard%
, Young%
\BCBL {}\ \BBA {} Brogan%
}{%
Draelos%
\ \protect \BOthers {.}}{%
{\protect \APACyear {2015}}%
}]{%
draelos2015new}
\APACinsertmetastar {%
draelos2015new}%
\begin{APACrefauthors}%
Draelos, T\BPBI J.%
, Ballard, S.%
, Young, C\BPBI J.%
\BCBL {}\ \BBA {} Brogan, R.%
\end{APACrefauthors}%
\unskip\
\newblock
\APACrefYearMonthDay{2015}{}{}.
\newblock
{\BBOQ}\APACrefatitle {A new method for producing automated seismic bulletins:
  Probabilistic event detection, association, and location} {A new method for
  producing automated seismic bulletins: Probabilistic event detection,
  association, and location}.{\BBCQ}
\newblock
\APACjournalVolNumPages{Bulletin of the Seismological Society of
  America}{105}{5}{2453--2467}.
\PrintBackRefs{\CurrentBib}

\bibitem [\protect \citeauthoryear {%
Ferguson%
}{%
Ferguson%
}{%
{\protect \APACyear {1973}}%
}]{%
ferguson1973bayesian}
\APACinsertmetastar {%
ferguson1973bayesian}%
\begin{APACrefauthors}%
Ferguson, T\BPBI S.%
\end{APACrefauthors}%
\unskip\
\newblock
\APACrefYearMonthDay{1973}{}{}.
\newblock
{\BBOQ}\APACrefatitle {A Bayesian analysis of some nonparametric problems} {A
  bayesian analysis of some nonparametric problems}.{\BBCQ}
\newblock
\APACjournalVolNumPages{The annals of statistics}{}{}{209--230}.
\PrintBackRefs{\CurrentBib}

\bibitem [\protect \citeauthoryear {%
Fletcher%
}{%
Fletcher%
}{%
{\protect \APACyear {2013}}%
}]{%
fletcher2013practical}
\APACinsertmetastar {%
fletcher2013practical}%
\begin{APACrefauthors}%
Fletcher, R.%
\end{APACrefauthors}%
\unskip\
\newblock
\APACrefYear{2013}.
\newblock
\APACrefbtitle {Practical methods of optimization} {Practical methods of
  optimization}.
\newblock
\APACaddressPublisher{}{John Wiley \& Sons}.
\PrintBackRefs{\CurrentBib}

\bibitem [\protect \citeauthoryear {%
Friberg%
, Lisowski%
, Dricker%
\BCBL {}\ \BBA {} Hellman%
}{%
Friberg%
\ \protect \BOthers {.}}{%
{\protect \APACyear {2010}}%
}]{%
friberg2010earthworm}
\APACinsertmetastar {%
friberg2010earthworm}%
\begin{APACrefauthors}%
Friberg, P.%
, Lisowski, S.%
, Dricker, I.%
\BCBL {}\ \BBA {} Hellman, S.%
\end{APACrefauthors}%
\unskip\
\newblock
\APACrefYearMonthDay{2010}{}{}.
\newblock
{\BBOQ}\APACrefatitle {Earthworm in the 21st century} {Earthworm in the 21st
  century}.{\BBCQ}
\newblock
\BIn{} \APACrefbtitle {EGU General Assembly Conference Abstracts} {Egu general
  assembly conference abstracts}\ (\BPG~12654).
\PrintBackRefs{\CurrentBib}

\bibitem [\protect \citeauthoryear {%
Gibbons%
, Kv{\ae}rna%
, Harris%
\BCBL {}\ \BBA {} Dodge%
}{%
Gibbons%
\ \protect \BOthers {.}}{%
{\protect \APACyear {2016}}%
}]{%
gibbons2016iterative}
\APACinsertmetastar {%
gibbons2016iterative}%
\begin{APACrefauthors}%
Gibbons, S\BPBI J.%
, Kv{\ae}rna, T.%
, Harris, D\BPBI B.%
\BCBL {}\ \BBA {} Dodge, D\BPBI A.%
\end{APACrefauthors}%
\unskip\
\newblock
\APACrefYearMonthDay{2016}{}{}.
\newblock
{\BBOQ}\APACrefatitle {Iterative strategies for aftershock classification in
  automatic seismic processing pipelines} {Iterative strategies for aftershock
  classification in automatic seismic processing pipelines}.{\BBCQ}
\newblock
\APACjournalVolNumPages{Seismological Research Letters}{87}{4}{919--929}.
\PrintBackRefs{\CurrentBib}

\bibitem [\protect \citeauthoryear {%
Gibbons%
\ \BBA {} Ringdal%
}{%
Gibbons%
\ \BBA {} Ringdal%
}{%
{\protect \APACyear {2006}}%
}]{%
gibbons2006detection}
\APACinsertmetastar {%
gibbons2006detection}%
\begin{APACrefauthors}%
Gibbons, S\BPBI J.%
\BCBT {}\ \BBA {} Ringdal, F.%
\end{APACrefauthors}%
\unskip\
\newblock
\APACrefYearMonthDay{2006}{}{}.
\newblock
{\BBOQ}\APACrefatitle {The detection of low magnitude seismic events using
  array-based waveform correlation} {The detection of low magnitude seismic
  events using array-based waveform correlation}.{\BBCQ}
\newblock
\APACjournalVolNumPages{Geophysical Journal International}{165}{1}{149--166}.
\PrintBackRefs{\CurrentBib}

\bibitem [\protect \citeauthoryear {%
Gutenberg%
}{%
Gutenberg%
}{%
{\protect \APACyear {1956}}%
}]{%
gutenberg1956energy}
\APACinsertmetastar {%
gutenberg1956energy}%
\begin{APACrefauthors}%
Gutenberg, B.%
\end{APACrefauthors}%
\unskip\
\newblock
\APACrefYearMonthDay{1956}{}{}.
\newblock
{\BBOQ}\APACrefatitle {The energy of earthquakes} {The energy of
  earthquakes}.{\BBCQ}
\newblock
\APACjournalVolNumPages{Quarterly Journal of the Geological
  Society}{112}{1-4}{1--14}.
\PrintBackRefs{\CurrentBib}

\bibitem [\protect \citeauthoryear {%
Harris%
\ \BBA {} Dodge%
}{%
Harris%
\ \BBA {} Dodge%
}{%
{\protect \APACyear {2011}}%
}]{%
harris2011}
\APACinsertmetastar {%
harris2011}%
\begin{APACrefauthors}%
Harris, D\BPBI B.%
\BCBT {}\ \BBA {} Dodge, D\BPBI A.%
\end{APACrefauthors}%
\unskip\
\newblock
\APACrefYearMonthDay{2011}{04}{}.
\newblock
{\BBOQ}\APACrefatitle {{An Autonomous System for Grouping Events in a
  Developing Aftershock Sequence}} {{An Autonomous System for Grouping Events
  in a Developing Aftershock Sequence}}.{\BBCQ}
\newblock
\APACjournalVolNumPages{Bulletin of the Seismological Society of
  America}{101}{2}{763-774}.
\newblock
\begin{APACrefDOI} \doi{10.1785/0120100103} \end{APACrefDOI}
\PrintBackRefs{\CurrentBib}

\bibitem [\protect \citeauthoryear {%
Hauksson%
, Yang%
\BCBL {}\ \BBA {} Shearer%
}{%
Hauksson%
\ \protect \BOthers {.}}{%
{\protect \APACyear {2012}}%
}]{%
hauksson2012waveform}
\APACinsertmetastar {%
hauksson2012waveform}%
\begin{APACrefauthors}%
Hauksson, E.%
, Yang, W.%
\BCBL {}\ \BBA {} Shearer, P\BPBI M.%
\end{APACrefauthors}%
\unskip\
\newblock
\APACrefYearMonthDay{2012}{}{}.
\newblock
{\BBOQ}\APACrefatitle {Waveform relocated earthquake catalog for southern
  California (1981 to June 2011)} {Waveform relocated earthquake catalog for
  southern california (1981 to june 2011)}.{\BBCQ}
\newblock
\APACjournalVolNumPages{Bulletin of the Seismological Society of
  America}{102}{5}{2239--2244}.
\PrintBackRefs{\CurrentBib}

\bibitem [\protect \citeauthoryear {%
Huber%
}{%
Huber%
}{%
{\protect \APACyear {1992}}%
}]{%
huber1992robust}
\APACinsertmetastar {%
huber1992robust}%
\begin{APACrefauthors}%
Huber, P\BPBI J.%
\end{APACrefauthors}%
\unskip\
\newblock
\APACrefYearMonthDay{1992}{}{}.
\newblock
{\BBOQ}\APACrefatitle {Robust estimation of a location parameter} {Robust
  estimation of a location parameter}.{\BBCQ}
\newblock
\BIn{} \APACrefbtitle {Breakthroughs in statistics} {Breakthroughs in
  statistics}\ (\BPGS\ 492--518).
\newblock
\APACaddressPublisher{}{Springer}.
\PrintBackRefs{\CurrentBib}

\bibitem [\protect \citeauthoryear {%
Klein%
}{%
Klein%
}{%
{\protect \APACyear {2002}}%
}]{%
klein2002user}
\APACinsertmetastar {%
klein2002user}%
\begin{APACrefauthors}%
Klein, F\BPBI W.%
\end{APACrefauthors}%
\unskip\
\newblock
\APACrefYearMonthDay{2002}{}{}.
\newblock
\APACrefbtitle {User's guide to HYPOINVERSE-2000, a Fortran program to solve
  for earthquake locations and magnitudes} {User's guide to hypoinverse-2000, a
  fortran program to solve for earthquake locations and magnitudes}\
  \APACbVolEdTR{}{\BTR{}}.
\newblock
\APACaddressInstitution{}{US Geological Survey}.
\PrintBackRefs{\CurrentBib}

\bibitem [\protect \citeauthoryear {%
Liu%
, Zhang%
, Zhu%
, Ellsworth%
\BCBL {}\ \BBA {} Li%
}{%
Liu%
\ \protect \BOthers {.}}{%
{\protect \APACyear {2020}}%
}]{%
liu2020rapid}
\APACinsertmetastar {%
liu2020rapid}%
\begin{APACrefauthors}%
Liu, M.%
, Zhang, M.%
, Zhu, W.%
, Ellsworth, W\BPBI L.%
\BCBL {}\ \BBA {} Li, H.%
\end{APACrefauthors}%
\unskip\
\newblock
\APACrefYearMonthDay{2020}{}{}.
\newblock
{\BBOQ}\APACrefatitle {Rapid characterization of the July 2019 Ridgecrest,
  California, earthquake sequence from raw seismic data using machine-learning
  phase picker} {Rapid characterization of the july 2019 ridgecrest,
  california, earthquake sequence from raw seismic data using machine-learning
  phase picker}.{\BBCQ}
\newblock
\APACjournalVolNumPages{Geophysical Research Letters}{47}{4}{e2019GL086189}.
\PrintBackRefs{\CurrentBib}

\bibitem [\protect \citeauthoryear {%
McBrearty%
, Delorey%
\BCBL {}\ \BBA {} Johnson%
}{%
McBrearty%
, Delorey%
\BCBL {}\ \BBA {} Johnson%
}{%
{\protect \APACyear {2019}}%
}]{%
mcbrearty2019pairwise}
\APACinsertmetastar {%
mcbrearty2019pairwise}%
\begin{APACrefauthors}%
McBrearty, I\BPBI W.%
, Delorey, A\BPBI A.%
\BCBL {}\ \BBA {} Johnson, P\BPBI A.%
\end{APACrefauthors}%
\unskip\
\newblock
\APACrefYearMonthDay{2019}{}{}.
\newblock
{\BBOQ}\APACrefatitle {Pairwise association of seismic arrivals with
  convolutional neural networks} {Pairwise association of seismic arrivals with
  convolutional neural networks}.{\BBCQ}
\newblock
\APACjournalVolNumPages{Seismological Research Letters}{90}{2A}{503--509}.
\PrintBackRefs{\CurrentBib}

\bibitem [\protect \citeauthoryear {%
McBrearty%
, Gomberg%
, Delorey%
\BCBL {}\ \BBA {} Johnson%
}{%
McBrearty%
, Gomberg%
\BCBL {}\ \protect \BOthers {.}}{%
{\protect \APACyear {2019}}%
}]{%
mcbrearty2019earthquake}
\APACinsertmetastar {%
mcbrearty2019earthquake}%
\begin{APACrefauthors}%
McBrearty, I\BPBI W.%
, Gomberg, J.%
, Delorey, A\BPBI A.%
\BCBL {}\ \BBA {} Johnson, P\BPBI A.%
\end{APACrefauthors}%
\unskip\
\newblock
\APACrefYearMonthDay{2019}{}{}.
\newblock
{\BBOQ}\APACrefatitle {Earthquake arrival association with backprojection and
  graph theory} {Earthquake arrival association with backprojection and graph
  theory}.{\BBCQ}
\newblock
\APACjournalVolNumPages{Bulletin of the Seismological Society of
  America}{109}{6}{2510--2531}.
\PrintBackRefs{\CurrentBib}

\bibitem [\protect \citeauthoryear {%
Melchior%
\ \BBA {} Goulding%
}{%
Melchior%
\ \BBA {} Goulding%
}{%
{\protect \APACyear {2018}}%
}]{%
melchior2018filling}
\APACinsertmetastar {%
melchior2018filling}%
\begin{APACrefauthors}%
Melchior, P.%
\BCBT {}\ \BBA {} Goulding, A\BPBI D.%
\end{APACrefauthors}%
\unskip\
\newblock
\APACrefYearMonthDay{2018}{{\APACmonth{10}}}{}.
\newblock
{\BBOQ}\APACrefatitle {Filling the Gaps: {{Gaussian}} Mixture Models from
  Noisy, Truncated or Incomplete Samples} {Filling the gaps: {{Gaussian}}
  mixture models from noisy, truncated or incomplete samples}.{\BBCQ}
\newblock
\APACjournalVolNumPages{Astronomy and Computing}{25}{}{183--194}.
\PrintBackRefs{\CurrentBib}

\bibitem [\protect \citeauthoryear {%
Mousavi%
, Ellsworth%
, Zhu%
, Chuang%
\BCBL {}\ \BBA {} Beroza%
}{%
Mousavi%
\ \protect \BOthers {.}}{%
{\protect \APACyear {2020}}%
}]{%
mousavi2020earthquake}
\APACinsertmetastar {%
mousavi2020earthquake}%
\begin{APACrefauthors}%
Mousavi, S\BPBI M.%
, Ellsworth, W\BPBI L.%
, Zhu, W.%
, Chuang, L\BPBI Y.%
\BCBL {}\ \BBA {} Beroza, G\BPBI C.%
\end{APACrefauthors}%
\unskip\
\newblock
\APACrefYearMonthDay{2020}{}{}.
\newblock
{\BBOQ}\APACrefatitle {Earthquake transformer—an attentive deep-learning
  model for simultaneous earthquake detection and phase picking} {Earthquake
  transformer—an attentive deep-learning model for simultaneous earthquake
  detection and phase picking}.{\BBCQ}
\newblock
\APACjournalVolNumPages{Nature communications}{11}{1}{1--12}.
\PrintBackRefs{\CurrentBib}

\bibitem [\protect \citeauthoryear {%
M{\"u}nchmeyer%
, Bindi%
, Sippl%
, Leser%
\BCBL {}\ \BBA {} Tilmann%
}{%
M{\"u}nchmeyer%
\ \protect \BOthers {.}}{%
{\protect \APACyear {2020}}%
}]{%
munchmeyer2020low}
\APACinsertmetastar {%
munchmeyer2020low}%
\begin{APACrefauthors}%
M{\"u}nchmeyer, J.%
, Bindi, D.%
, Sippl, C.%
, Leser, U.%
\BCBL {}\ \BBA {} Tilmann, F.%
\end{APACrefauthors}%
\unskip\
\newblock
\APACrefYearMonthDay{2020}{}{}.
\newblock
{\BBOQ}\APACrefatitle {Low uncertainty multifeature magnitude estimation with
  3-D corrections and boosting tree regression: application to North Chile}
  {Low uncertainty multifeature magnitude estimation with 3-d corrections and
  boosting tree regression: application to north chile}.{\BBCQ}
\newblock
\APACjournalVolNumPages{Geophysical Journal International}{220}{1}{142--159}.
\PrintBackRefs{\CurrentBib}

\bibitem [\protect \citeauthoryear {%
Park%
, Mousavi%
, Zhu%
, Ellsworth%
\BCBL {}\ \BBA {} Beroza%
}{%
Park%
\ \protect \BOthers {.}}{%
{\protect \APACyear {2020}}%
}]{%
park2020machine}
\APACinsertmetastar {%
park2020machine}%
\begin{APACrefauthors}%
Park, Y.%
, Mousavi, S\BPBI M.%
, Zhu, W.%
, Ellsworth, W\BPBI L.%
\BCBL {}\ \BBA {} Beroza, G\BPBI C.%
\end{APACrefauthors}%
\unskip\
\newblock
\APACrefYearMonthDay{2020}{}{}.
\newblock
{\BBOQ}\APACrefatitle {Machine-learning-based analysis of the Guy-Greenbrier,
  Arkansas earthquakes: A tale of two sequences} {Machine-learning-based
  analysis of the guy-greenbrier, arkansas earthquakes: A tale of two
  sequences}.{\BBCQ}
\newblock
\APACjournalVolNumPages{Geophysical Research Letters}{47}{6}{e2020GL087032}.
\PrintBackRefs{\CurrentBib}

\bibitem [\protect \citeauthoryear {%
Peng%
\ \BBA {} Zhao%
}{%
Peng%
\ \BBA {} Zhao%
}{%
{\protect \APACyear {2009}}%
}]{%
peng2009migration}
\APACinsertmetastar {%
peng2009migration}%
\begin{APACrefauthors}%
Peng, Z.%
\BCBT {}\ \BBA {} Zhao, P.%
\end{APACrefauthors}%
\unskip\
\newblock
\APACrefYearMonthDay{2009}{}{}.
\newblock
{\BBOQ}\APACrefatitle {Migration of early aftershocks following the 2004
  Parkfield earthquake} {Migration of early aftershocks following the 2004
  parkfield earthquake}.{\BBCQ}
\newblock
\APACjournalVolNumPages{Nature Geoscience}{2}{12}{877--881}.
\PrintBackRefs{\CurrentBib}

\bibitem [\protect \citeauthoryear {%
Permuter%
, Francos%
\BCBL {}\ \BBA {} Jermyn%
}{%
Permuter%
\ \protect \BOthers {.}}{%
{\protect \APACyear {2006}}%
}]{%
permuter2006study}
\APACinsertmetastar {%
permuter2006study}%
\begin{APACrefauthors}%
Permuter, H.%
, Francos, J.%
\BCBL {}\ \BBA {} Jermyn, I.%
\end{APACrefauthors}%
\unskip\
\newblock
\APACrefYearMonthDay{2006}{}{}.
\newblock
{\BBOQ}\APACrefatitle {A study of Gaussian mixture models of color and texture
  features for image classification and segmentation} {A study of gaussian
  mixture models of color and texture features for image classification and
  segmentation}.{\BBCQ}
\newblock
\APACjournalVolNumPages{Pattern Recognition}{39}{4}{695--706}.
\PrintBackRefs{\CurrentBib}

\bibitem [\protect \citeauthoryear {%
Picozzi%
, Bindi%
, Spallarossa%
, Di~Giacomo%
\BCBL {}\ \BBA {} Zollo%
}{%
Picozzi%
\ \protect \BOthers {.}}{%
{\protect \APACyear {2018}}%
}]{%
picozzi2018rapid}
\APACinsertmetastar {%
picozzi2018rapid}%
\begin{APACrefauthors}%
Picozzi, M.%
, Bindi, D.%
, Spallarossa, D.%
, Di~Giacomo, D.%
\BCBL {}\ \BBA {} Zollo, A.%
\end{APACrefauthors}%
\unskip\
\newblock
\APACrefYearMonthDay{2018}{{\APACmonth{06}}}{}.
\newblock
{\BBOQ}\APACrefatitle {A Rapid Response Magnitude Scale for Timely Assessment
  of the High Frequency Seismic Radiation} {A rapid response magnitude scale
  for timely assessment of the high frequency seismic radiation}.{\BBCQ}
\newblock
\APACjournalVolNumPages{Scientific Reports}{8}{1}{8562}.
\PrintBackRefs{\CurrentBib}

\bibitem [\protect \citeauthoryear {%
Reynolds%
\ \BBA {} Rose%
}{%
Reynolds%
\ \BBA {} Rose%
}{%
{\protect \APACyear {1995}}%
}]{%
reynolds1995robust}
\APACinsertmetastar {%
reynolds1995robust}%
\begin{APACrefauthors}%
Reynolds, D\BPBI A.%
\BCBT {}\ \BBA {} Rose, R\BPBI C.%
\end{APACrefauthors}%
\unskip\
\newblock
\APACrefYearMonthDay{1995}{}{}.
\newblock
{\BBOQ}\APACrefatitle {Robust text-independent speaker identification using
  Gaussian mixture speaker models} {Robust text-independent speaker
  identification using gaussian mixture speaker models}.{\BBCQ}
\newblock
\APACjournalVolNumPages{IEEE transactions on speech and audio
  processing}{3}{1}{72--83}.
\PrintBackRefs{\CurrentBib}

\bibitem [\protect \citeauthoryear {%
Richter%
}{%
Richter%
}{%
{\protect \APACyear {1935}}%
}]{%
richter1935instrumental}
\APACinsertmetastar {%
richter1935instrumental}%
\begin{APACrefauthors}%
Richter, C\BPBI F.%
\end{APACrefauthors}%
\unskip\
\newblock
\APACrefYearMonthDay{1935}{}{}.
\newblock
{\BBOQ}\APACrefatitle {An instrumental earthquake magnitude scale} {An
  instrumental earthquake magnitude scale}.{\BBCQ}
\newblock
\APACjournalVolNumPages{Bulletin of the seismological society of
  America}{25}{1}{1--32}.
\PrintBackRefs{\CurrentBib}

\bibitem [\protect \citeauthoryear {%
Ross%
, Idini%
\BCBL {}\ \protect \BOthers {.}}{%
Ross%
, Idini%
\BCBL {}\ \protect \BOthers {.}}{%
{\protect \APACyear {2019}}%
}]{%
ross2019hierarchical}
\APACinsertmetastar {%
ross2019hierarchical}%
\begin{APACrefauthors}%
Ross, Z\BPBI E.%
, Idini, B.%
, Jia, Z.%
, Stephenson, O\BPBI L.%
, Zhong, M.%
, Wang, X.%
\BDBL {}others%
\end{APACrefauthors}%
\unskip\
\newblock
\APACrefYearMonthDay{2019}{}{}.
\newblock
{\BBOQ}\APACrefatitle {Hierarchical interlocked orthogonal faulting in the 2019
  Ridgecrest earthquake sequence} {Hierarchical interlocked orthogonal faulting
  in the 2019 ridgecrest earthquake sequence}.{\BBCQ}
\newblock
\APACjournalVolNumPages{Science}{366}{6463}{346--351}.
\PrintBackRefs{\CurrentBib}

\bibitem [\protect \citeauthoryear {%
Ross%
, Meier%
, Hauksson%
\BCBL {}\ \BBA {} Heaton%
}{%
Ross%
\ \protect \BOthers {.}}{%
{\protect \APACyear {2018}}%
}]{%
ross2018generalized}
\APACinsertmetastar {%
ross2018generalized}%
\begin{APACrefauthors}%
Ross, Z\BPBI E.%
, Meier, M\BHBI A.%
, Hauksson, E.%
\BCBL {}\ \BBA {} Heaton, T\BPBI H.%
\end{APACrefauthors}%
\unskip\
\newblock
\APACrefYearMonthDay{2018}{}{}.
\newblock
{\BBOQ}\APACrefatitle {Generalized seismic phase detection with deep learning}
  {Generalized seismic phase detection with deep learning}.{\BBCQ}
\newblock
\APACjournalVolNumPages{Bulletin of the Seismological Society of
  America}{108}{5A}{2894--2901}.
\PrintBackRefs{\CurrentBib}

\bibitem [\protect \citeauthoryear {%
Ross%
, Trugman%
, Azizzadenesheli%
\BCBL {}\ \BBA {} Anandkumar%
}{%
Ross%
\ \protect \BOthers {.}}{%
{\protect \APACyear {2020}}%
}]{%
ross2020directivity}
\APACinsertmetastar {%
ross2020directivity}%
\begin{APACrefauthors}%
Ross, Z\BPBI E.%
, Trugman, D\BPBI T.%
, Azizzadenesheli, K.%
\BCBL {}\ \BBA {} Anandkumar, A.%
\end{APACrefauthors}%
\unskip\
\newblock
\APACrefYearMonthDay{2020}{}{}.
\newblock
{\BBOQ}\APACrefatitle {Directivity modes of earthquake populations with
  unsupervised learning} {Directivity modes of earthquake populations with
  unsupervised learning}.{\BBCQ}
\newblock
\APACjournalVolNumPages{Journal of Geophysical Research: Solid
  Earth}{125}{2}{e2019JB018299}.
\PrintBackRefs{\CurrentBib}

\bibitem [\protect \citeauthoryear {%
Ross%
, Trugman%
, Hauksson%
\BCBL {}\ \BBA {} Shearer%
}{%
Ross%
, Trugman%
\BCBL {}\ \protect \BOthers {.}}{%
{\protect \APACyear {2019}}%
}]{%
ross2019searching}
\APACinsertmetastar {%
ross2019searching}%
\begin{APACrefauthors}%
Ross, Z\BPBI E.%
, Trugman, D\BPBI T.%
, Hauksson, E.%
\BCBL {}\ \BBA {} Shearer, P\BPBI M.%
\end{APACrefauthors}%
\unskip\
\newblock
\APACrefYearMonthDay{2019}{}{}.
\newblock
{\BBOQ}\APACrefatitle {Searching for hidden earthquakes in Southern California}
  {Searching for hidden earthquakes in southern california}.{\BBCQ}
\newblock
\APACjournalVolNumPages{Science}{364}{6442}{767--771}.
\PrintBackRefs{\CurrentBib}

\bibitem [\protect \citeauthoryear {%
Ross%
, Yue%
, Meier%
, Hauksson%
\BCBL {}\ \BBA {} Heaton%
}{%
Ross%
, Yue%
\BCBL {}\ \protect \BOthers {.}}{%
{\protect \APACyear {2019}}%
}]{%
ross2019phaselink}
\APACinsertmetastar {%
ross2019phaselink}%
\begin{APACrefauthors}%
Ross, Z\BPBI E.%
, Yue, Y.%
, Meier, M\BHBI A.%
, Hauksson, E.%
\BCBL {}\ \BBA {} Heaton, T\BPBI H.%
\end{APACrefauthors}%
\unskip\
\newblock
\APACrefYearMonthDay{2019}{}{}.
\newblock
{\BBOQ}\APACrefatitle {PhaseLink: A deep learning approach to seismic phase
  association} {Phaselink: A deep learning approach to seismic phase
  association}.{\BBCQ}
\newblock
\APACjournalVolNumPages{Journal of Geophysical Research: Solid
  Earth}{124}{1}{856--869}.
\PrintBackRefs{\CurrentBib}

\bibitem [\protect \citeauthoryear {%
SCEDC%
}{%
SCEDC%
}{%
{\protect \APACyear {2013}}%
}]{%
southern2013southern}
\APACinsertmetastar {%
southern2013southern}%
\begin{APACrefauthors}%
SCEDC.%
\end{APACrefauthors}%
\unskip\
\newblock
\APACrefYearMonthDay{2013}{}{}.
\newblock
{\BBOQ}\APACrefatitle {{Southern California Earthquake Center}} {{Southern
  California Earthquake Center}}.{\BBCQ}
\newblock
\APACjournalVolNumPages{Caltech. Dataset}{}{}{}.
\PrintBackRefs{\CurrentBib}

\bibitem [\protect \citeauthoryear {%
Schubert%
, Sander%
, Ester%
, Kriegel%
\BCBL {}\ \BBA {} Xu%
}{%
Schubert%
\ \protect \BOthers {.}}{%
{\protect \APACyear {2017}}%
}]{%
schubert2017dbscan}
\APACinsertmetastar {%
schubert2017dbscan}%
\begin{APACrefauthors}%
Schubert, E.%
, Sander, J.%
, Ester, M.%
, Kriegel, H\BPBI P.%
\BCBL {}\ \BBA {} Xu, X.%
\end{APACrefauthors}%
\unskip\
\newblock
\APACrefYearMonthDay{2017}{}{}.
\newblock
{\BBOQ}\APACrefatitle {DBSCAN revisited, revisited: why and how you should
  (still) use DBSCAN} {Dbscan revisited, revisited: why and how you should
  (still) use dbscan}.{\BBCQ}
\newblock
\APACjournalVolNumPages{ACM Transactions on Database Systems
  (TODS)}{42}{3}{1--21}.
\PrintBackRefs{\CurrentBib}

\bibitem [\protect \citeauthoryear {%
Seydoux%
\ \protect \BOthers {.}}{%
Seydoux%
\ \protect \BOthers {.}}{%
{\protect \APACyear {2020}}%
}]{%
seydoux2020clustering}
\APACinsertmetastar {%
seydoux2020clustering}%
\begin{APACrefauthors}%
Seydoux, L.%
, Balestriero, R.%
, Poli, P.%
, De~Hoop, M.%
, Campillo, M.%
\BCBL {}\ \BBA {} Baraniuk, R.%
\end{APACrefauthors}%
\unskip\
\newblock
\APACrefYearMonthDay{2020}{}{}.
\newblock
{\BBOQ}\APACrefatitle {Clustering earthquake signals and background noises in
  continuous seismic data with unsupervised deep learning} {Clustering
  earthquake signals and background noises in continuous seismic data with
  unsupervised deep learning}.{\BBCQ}
\newblock
\APACjournalVolNumPages{Nature communications}{11}{1}{1--12}.
\PrintBackRefs{\CurrentBib}

\bibitem [\protect \citeauthoryear {%
Shelly%
}{%
Shelly%
}{%
{\protect \APACyear {2020}}%
}]{%
shelly2020high}
\APACinsertmetastar {%
shelly2020high}%
\begin{APACrefauthors}%
Shelly, D\BPBI R.%
\end{APACrefauthors}%
\unskip\
\newblock
\APACrefYearMonthDay{2020}{}{}.
\newblock
{\BBOQ}\APACrefatitle {A high-resolution seismic catalog for the initial 2019
  Ridgecrest earthquake sequence: Foreshocks, aftershocks, and faulting
  complexity} {A high-resolution seismic catalog for the initial 2019
  ridgecrest earthquake sequence: Foreshocks, aftershocks, and faulting
  complexity}.{\BBCQ}
\newblock
\APACjournalVolNumPages{Seismological Research Letters}{91}{4}{1971--1978}.
\PrintBackRefs{\CurrentBib}

\bibitem [\protect \citeauthoryear {%
Shelly%
, Beroza%
\BCBL {}\ \BBA {} Ide%
}{%
Shelly%
\ \protect \BOthers {.}}{%
{\protect \APACyear {2007}}%
}]{%
shelly2007non}
\APACinsertmetastar {%
shelly2007non}%
\begin{APACrefauthors}%
Shelly, D\BPBI R.%
, Beroza, G\BPBI C.%
\BCBL {}\ \BBA {} Ide, S.%
\end{APACrefauthors}%
\unskip\
\newblock
\APACrefYearMonthDay{2007}{}{}.
\newblock
{\BBOQ}\APACrefatitle {Non-volcanic tremor and low-frequency earthquake swarms}
  {Non-volcanic tremor and low-frequency earthquake swarms}.{\BBCQ}
\newblock
\APACjournalVolNumPages{Nature}{446}{7133}{305--307}.
\PrintBackRefs{\CurrentBib}

\bibitem [\protect \citeauthoryear {%
Tan%
\ \protect \BOthers {.}}{%
Tan%
\ \protect \BOthers {.}}{%
{\protect \APACyear {2021}}%
}]{%
tan2021machine}
\APACinsertmetastar {%
tan2021machine}%
\begin{APACrefauthors}%
Tan, Y\BPBI J.%
, Waldhauser, F.%
, Ellsworth, W\BPBI L.%
, Zhang, M.%
, Zhu, W.%
, Michele, M.%
\BDBL {}Segou, M.%
\end{APACrefauthors}%
\unskip\
\newblock
\APACrefYearMonthDay{2021}{}{}.
\newblock
{\BBOQ}\APACrefatitle {Machine-Learning-Based High-Resolution Earthquake
  Catalog Reveals How Complex Fault Structures Were Activated during the
  2016--2017 Central Italy Sequence} {Machine-learning-based high-resolution
  earthquake catalog reveals how complex fault structures were activated during
  the 2016--2017 central italy sequence}.{\BBCQ}
\newblock
\APACjournalVolNumPages{The Seismic Record}{1}{1}{11--19}.
\PrintBackRefs{\CurrentBib}

\bibitem [\protect \citeauthoryear {%
Waldhauser%
\ \BBA {} Schaff%
}{%
Waldhauser%
\ \BBA {} Schaff%
}{%
{\protect \APACyear {2008}}%
}]{%
waldhauser2008large}
\APACinsertmetastar {%
waldhauser2008large}%
\begin{APACrefauthors}%
Waldhauser, F.%
\BCBT {}\ \BBA {} Schaff, D\BPBI P.%
\end{APACrefauthors}%
\unskip\
\newblock
\APACrefYearMonthDay{2008}{}{}.
\newblock
{\BBOQ}\APACrefatitle {Large-scale relocation of two decades of Northern
  California seismicity using cross-correlation and double-difference methods}
  {Large-scale relocation of two decades of northern california seismicity
  using cross-correlation and double-difference methods}.{\BBCQ}
\newblock
\APACjournalVolNumPages{Journal of Geophysical Research: Solid
  Earth}{113}{B8}{}.
\PrintBackRefs{\CurrentBib}

\bibitem [\protect \citeauthoryear {%
Weber%
\ \protect \BOthers {.}}{%
Weber%
\ \protect \BOthers {.}}{%
{\protect \APACyear {2007}}%
}]{%
weber2007seiscomp3}
\APACinsertmetastar {%
weber2007seiscomp3}%
\begin{APACrefauthors}%
Weber, B.%
, Becker, J.%
, Hanka, W.%
, Heinloo, A.%
, Hoffmann, M.%
, Kraft, T.%
\BDBL {}Thoms, H.%
\end{APACrefauthors}%
\unskip\
\newblock
\APACrefYearMonthDay{2007}{}{}.
\newblock
{\BBOQ}\APACrefatitle {SeisComP3—Automatic and interactive real time data
  processing} {Seiscomp3—automatic and interactive real time data
  processing}.{\BBCQ}
\newblock
\BIn{} \APACrefbtitle {Geophysical Research Abstracts} {Geophysical research
  abstracts}\ (\BVOL~9).
\PrintBackRefs{\CurrentBib}

\bibitem [\protect \citeauthoryear {%
Wishart%
}{%
Wishart%
}{%
{\protect \APACyear {1928}}%
}]{%
wishart1928generalised}
\APACinsertmetastar {%
wishart1928generalised}%
\begin{APACrefauthors}%
Wishart, J.%
\end{APACrefauthors}%
\unskip\
\newblock
\APACrefYearMonthDay{1928}{}{}.
\newblock
{\BBOQ}\APACrefatitle {The generalised product moment distribution in samples
  from a normal multivariate population} {The generalised product moment
  distribution in samples from a normal multivariate population}.{\BBCQ}
\newblock
\APACjournalVolNumPages{Biometrika}{}{}{32--52}.
\PrintBackRefs{\CurrentBib}

\bibitem [\protect \citeauthoryear {%
Woollam%
, Rietbrock%
, Leitloff%
\BCBL {}\ \BBA {} Hinz%
}{%
Woollam%
\ \protect \BOthers {.}}{%
{\protect \APACyear {2020}}%
}]{%
woollam2020hex}
\APACinsertmetastar {%
woollam2020hex}%
\begin{APACrefauthors}%
Woollam, J.%
, Rietbrock, A.%
, Leitloff, J.%
\BCBL {}\ \BBA {} Hinz, S.%
\end{APACrefauthors}%
\unskip\
\newblock
\APACrefYearMonthDay{2020}{{\APACmonth{07}}}{}.
\newblock
{\BBOQ}\APACrefatitle {{{HEX}}: {{Hyperbolic Event eXtractor}}, a {{Seismic
  Phase Associator}} for {{Highly Active Seismic Regions}}} {{{HEX}}:
  {{Hyperbolic Event eXtractor}}, a {{Seismic Phase Associator}} for {{Highly
  Active Seismic Regions}}}.{\BBCQ}
\newblock
\APACjournalVolNumPages{Seismological Research Letters}{91}{5}{2769--2778}.
\PrintBackRefs{\CurrentBib}

\bibitem [\protect \citeauthoryear {%
Yeck%
\ \protect \BOthers {.}}{%
Yeck%
\ \protect \BOthers {.}}{%
{\protect \APACyear {2019}}%
}]{%
yeck2019glass3}
\APACinsertmetastar {%
yeck2019glass3}%
\begin{APACrefauthors}%
Yeck, W\BPBI L.%
, Patton, J\BPBI M.%
, Johnson, C\BPBI E.%
, Kragness, D.%
, Benz, H\BPBI M.%
, Earle, P\BPBI S.%
\BDBL {}Ambruz, N\BPBI B.%
\end{APACrefauthors}%
\unskip\
\newblock
\APACrefYearMonthDay{2019}{}{}.
\newblock
{\BBOQ}\APACrefatitle {GLASS3: A Standalone Multiscale Seismic Detection
  AssociatorGLASS3: A Standalone Multiscale Seismic Detection Associator}
  {Glass3: A standalone multiscale seismic detection associatorglass3: A
  standalone multiscale seismic detection associator}.{\BBCQ}
\newblock
\APACjournalVolNumPages{Bulletin of the Seismological Society of
  America}{109}{4}{1469--1478}.
\PrintBackRefs{\CurrentBib}

\bibitem [\protect \citeauthoryear {%
Yoon%
, O’Reilly%
, Bergen%
\BCBL {}\ \BBA {} Beroza%
}{%
Yoon%
\ \protect \BOthers {.}}{%
{\protect \APACyear {2015}}%
}]{%
yoon2015earthquake}
\APACinsertmetastar {%
yoon2015earthquake}%
\begin{APACrefauthors}%
Yoon, C\BPBI E.%
, O’Reilly, O.%
, Bergen, K\BPBI J.%
\BCBL {}\ \BBA {} Beroza, G\BPBI C.%
\end{APACrefauthors}%
\unskip\
\newblock
\APACrefYearMonthDay{2015}{}{}.
\newblock
{\BBOQ}\APACrefatitle {Earthquake detection through computationally efficient
  similarity search} {Earthquake detection through computationally efficient
  similarity search}.{\BBCQ}
\newblock
\APACjournalVolNumPages{Science advances}{1}{11}{e1501057}.
\PrintBackRefs{\CurrentBib}

\bibitem [\protect \citeauthoryear {%
Zhang%
, Ellsworth%
\BCBL {}\ \BBA {} Beroza%
}{%
Zhang%
\ \protect \BOthers {.}}{%
{\protect \APACyear {2019}}%
}]{%
zhang2019rapid}
\APACinsertmetastar {%
zhang2019rapid}%
\begin{APACrefauthors}%
Zhang, M.%
, Ellsworth, W\BPBI L.%
\BCBL {}\ \BBA {} Beroza, G\BPBI C.%
\end{APACrefauthors}%
\unskip\
\newblock
\APACrefYearMonthDay{2019}{}{}.
\newblock
{\BBOQ}\APACrefatitle {Rapid earthquake association and location} {Rapid
  earthquake association and location}.{\BBCQ}
\newblock
\APACjournalVolNumPages{Seismological Research Letters}{90}{6}{2276--2284}.
\PrintBackRefs{\CurrentBib}

\bibitem [\protect \citeauthoryear {%
L.~Zhu%
, Chuang%
, McClellan%
, Liu%
\BCBL {}\ \BBA {} Peng%
}{%
L.~Zhu%
\ \protect \BOthers {.}}{%
{\protect \APACyear {2021}}%
}]{%
zhu2021multi}
\APACinsertmetastar {%
zhu2021multi}%
\begin{APACrefauthors}%
Zhu, L.%
, Chuang, L.%
, McClellan, J\BPBI H.%
, Liu, E.%
\BCBL {}\ \BBA {} Peng, Z.%
\end{APACrefauthors}%
\unskip\
\newblock
\APACrefYearMonthDay{2021}{}{}.
\newblock
{\BBOQ}\APACrefatitle {A multi-channel approach for automatic microseismic
  event association using RANSAC-based arrival time event clustering (RATEC)}
  {A multi-channel approach for automatic microseismic event association using
  ransac-based arrival time event clustering (ratec)}.{\BBCQ}
\newblock
\APACjournalVolNumPages{Earthquake Research Advances}{}{}{100008}.
\PrintBackRefs{\CurrentBib}

\bibitem [\protect \citeauthoryear {%
W.~Zhu%
\ \BBA {} Beroza%
}{%
W.~Zhu%
\ \BBA {} Beroza%
}{%
{\protect \APACyear {2018}}%
}]{%
zhu2019phasenet}
\APACinsertmetastar {%
zhu2019phasenet}%
\begin{APACrefauthors}%
Zhu, W.%
\BCBT {}\ \BBA {} Beroza, G\BPBI C.%
\end{APACrefauthors}%
\unskip\
\newblock
\APACrefYearMonthDay{2018}{}{}.
\newblock
{\BBOQ}\APACrefatitle {PhaseNet: a deep-neural-network-based seismic
  arrival-time picking method} {Phasenet: a deep-neural-network-based seismic
  arrival-time picking method}.{\BBCQ}
\newblock
\APACjournalVolNumPages{Geophysical Journal International}{216}{1}{261--273}.
\PrintBackRefs{\CurrentBib}

\end{thebibliography}

\end{document}